\documentclass[conference]{IEEEtran}
\IEEEoverridecommandlockouts
\usepackage{cite}
\usepackage{amsmath,amssymb,amsfonts}
\usepackage{algorithmic}
\usepackage{graphicx}
\usepackage{textcomp}
\usepackage{xcolor}
\usepackage{hyperref}
\usepackage{tikz}
\usepackage{array}
\usepackage{wrapfig}
\usepackage{multirow}
\usepackage{tabularx}
\usepackage{makecell}
\usepackage[T1]{fontenc}
\usepackage{float}
\restylefloat{table}
\usepackage{subcaption}
\usepackage[ruled,vlined,linesnumbered]{algorithm2e}
\usepackage{pgfplotstable}
\usepackage{caption}

\usepackage{fancyvrb,lipsum}

\usepackage{url}

\def\BibTeX{{\rm B\kern-.05em{\sc i\kern-.025em b}\kern-.08em
    T\kern-.1667em\lower.7ex\hbox{E}\kern-.125emX}}

\newcolumntype{P}[1]{>{\centering\arraybackslash}p{#1}}

\usepackage{booktabs}
\usepackage{siunitx} 

\newcommand{\grover}{{GROVER}}
\newcommand{\gltr}{{GLTR}}
\newcommand{\gltrgpt}{{GLTR-GPT2}}
\newcommand{\gltrbert}{{GLTR-BERT}}
\newcommand{\gbert}{{BERT-Defense}}
\newcommand{\rbert}{{RoBERTa-Defense}}
\newcommand{\bert}{{BERT}}
\newcommand{\fast}{{FAST}}

\newcommand{\ourattack}{{DFTFooler}}
\newcommand{\gruen} {{GRN}}
\newcommand{\para}[1]{{\vspace{1pt} \bf \noindent #1 \hspace{6pt}}}

\newcommand{\eg}{{\em e.g.,\ }}
\newcommand{\ie}{{\em i.e.,\ }}
\newcommand{\etal}{{\em et al.}}

\usepackage{filehook}

\pgfplotsset{compat=1.17}

\begin{document}

\title{Deepfake Text Detection: Limitations and Opportunities}
\author{\IEEEauthorblockN{Jiameng Pu\IEEEauthorrefmark{1}\IEEEauthorrefmark{5},
Zain Sarwar\IEEEauthorrefmark{2}\IEEEauthorrefmark{5}\footnotetext{Indicates equal contribution},
Sifat Muhammad Abdullah\IEEEauthorrefmark{1}, 
Abdullah Rehman\IEEEauthorrefmark{1},
Yoonjin Kim\IEEEauthorrefmark{1},\\
Parantapa Bhattacharya\IEEEauthorrefmark{4},
Mobin Javed\IEEEauthorrefmark{3},
Bimal Viswanath\IEEEauthorrefmark{1}}

\IEEEauthorrefmark{1}Virginia Tech, \IEEEauthorrefmark{2}University of Chicago, \IEEEauthorrefmark{3}LUMS Pakistan,
\IEEEauthorrefmark{4}University of Virginia
\\
\IEEEauthorrefmark{1}\{jmpu, sifat, abdullahzr, ykim05, vbimal\}@vt.edu, \IEEEauthorrefmark{2}zsarwar@uchicago.edu, \IEEEauthorrefmark{4}parantapa@virginia.edu, \\\IEEEauthorrefmark{3}mobin.javed@lums.edu.pk  
}

\maketitle

\def\thefootnote{\IEEEauthorrefmark{5}}\footnotetext{These authors contributed equally to this work.}\def\thefootnote{\arabic{footnote}}

\begin{abstract}
Recent advances in generative models for language have enabled the creation of convincing synthetic text or deepfake text. Prior work has demonstrated the potential for misuse of deepfake text to mislead content consumers. Therefore, deepfake text detection, the task of discriminating between human and machine-generated text, is becoming increasingly critical. Several defenses have been proposed for deepfake text detection. However, we lack a thorough understanding of their real-world applicability. In this paper, we collect deepfake text from 4 online services powered by Transformer-based tools to evaluate the generalization ability of the defenses on content in the wild. We develop several low-cost adversarial attacks, and investigate the robustness of existing defenses against an adaptive attacker. We find that many defenses show significant degradation in performance under our evaluation scenarios compared to their original claimed performance. Our evaluation shows that tapping into the semantic information in the text content is a promising approach for improving the robustness and generalization performance of deepfake text detection schemes. 
\end{abstract}


    \begin{IEEEkeywords}
deepfake text, deepfake detection
\end{IEEEkeywords}

\section{Introduction}
\noindent 
Progress in natural language generation (NLG) has enabled deep learning models such as GPT-2~\cite{radford2019language} and GPT-3~\cite{brown2020language} to generate \textit{synthetic text} or \textit{deepfake text} with high linguistic quality. Both models fall in the Transformer~\cite{vaswani2017attention} family of language models (LMs). These large LMs with billions of parameters, can generate synthetic text on diverse topics. They have many 
applications, including, generating content for entertainment purposes (\eg stories, jokes)~\cite{xu2020megatron}, dialog systems~\cite{mi2020continual}, text summarization~\cite{zhang2019pretraining}, and automated journalism~\cite{leppanen2017data}. 

Unfortunately, such technology raises serious security concerns---synthetic text can be misused to power several threats aimed at misleading content consumers. Zellers \etal{} demonstrated that GPT-2-based LMs can generate convincing fake news articles~\cite{zellers2019defending}. Such tools can enable large-scale disinformation campaigns. Yao \etal{}~\cite{yao-2017-dnnreviews} showed that synthetic text can be used to create fake online restaurant reviews. Such threats will lead to users losing trust in online content, including crowd-sourced information. Other serious threats include automated email generation for targeted attacks~\cite{das2019automated}, and synthetic text that can radicalize individuals into having violent, extremist ideologies~\cite{mcguffie2020radicalization}. 

One approach to curb the misuse of synthetic text is to have humans evaluate and flag the text. However, such an approach is unlikely to scale, and more importantly, existing works already demonstrate that humans are unable to reliably distinguish between real and synthetic text. Recent work showed that the GPT-3~\cite{brown2020language} model can produce realistic news articles, which humans identify as synthetic only 52\% of the time. Therefore, it is pertinent that we develop robust automated schemes to accurately detect synthetic text. 

Fortunately, the research community has developed several defenses to automatically detect synthetic text~\cite{zellers2019defending, gehrmann2019gltr, ippolito2019human, zhong2020neural, yao-2017-dnnreviews}. They are all supervised learning schemes that use a language model to extract features to build a classifier. Some defenses only focus on features that capture statistical artifacts or imperfections in the generated text~\cite{gehrmann2019gltr}. Our reproduction of 6 of the best defenses show that they all achieve high detection performance, ranging from 79.6\% F1 score to 98.5\% F1 score in detecting synthetic samples. However, we lack a thorough understanding of the real-world applicability of these defenses.

To understand real-world applicability, we focus on two key aspects: (1) \textit{How well would existing defenses perform when applied to synthetic text in the wild?} All existing defenses have only been tested on synthetic datasets produced by the research community themselves. It is unclear how well they would work in the real world. To make matters more challenging, there is a lack of real-world synthetic datasets that the research community can use to study such defenses. (2) \textit{How would existing defenses perform against an adaptive attacker who is knowledgeable about the defenses?} Existing defense work provides limited understanding of threats posed by adaptive attackers. Of course, the community is well aware of attacks that craft adversarial samples to fool text classifiers~\cite{morris2020textattack}. Such attacks are also feasible against synthetic text detection classifiers. However, such adversarial attacks may not always be practical. For example, text adversarial attack schemes like TextFooler assume a black-box scenario that requires a large number of queries to the victim defense model to craft adversarial samples. Such query-based attacks could be caught by looking for specific query patterns~\cite{chen2020stateful, li2020blacklight, byun2021small}. Instead, we argue that there are much simpler, low-cost (computationally), yet effective strategies to fool existing defenses. To address these research questions, we conduct a study of synthetic text detection schemes through the lens of security. Our contributions are as follows:

\begin{itemize}
    \item We conduct a measurement study to collect synthetic text in the wild, and introduce 4 new real-world synthetic datasets. This includes synthetic data obtained from 3 \textit{text-generation-as-a-service} platforms (geared towards the SEO community), and synthetic data produced by a GPT-3 powered bot on the web (on Reddit.com). We find that Internet users and services are already using state-of-the-art Transformer-based models to synthesize text. 
    \item We evaluate performance of 6 state-of-the-art defenses on our new real-world synthetic datasets. Many defenses show significant degradation in performance compared to their original claimed/reproduced performance. Defenses using entity-based semantic features and those using robust pre-training methods combined with bidirectional context, generalize better to content in the wild.

    \item We develop simple, computationally low-cost attacks that modify the attacker's text generation process to evade detection. Our experiments show that just changing the decoding or text sampling strategy is sufficient to break many defenses. Moreover, defenses that are trained to look for specific text decoding artifacts are easier to evade by changing the text decoding strategy.
    \item We propose and evaluate a new black-box adversarial sample crafting strategy called \ourattack{}.
    Our attack requires no queries to the defense model, and exploits insights unique to the synthetic text detection problem. \ourattack{} only requires a publicly available pre-trained language model to craft adversarial perturbations. 
    \ourattack{} can produce transferable adversarial samples that can degrade the performance of multiple defenses.
    \item Lastly, our analysis shows that a promising approach to improve defenses is to tap into the semantic information in the text content. Our analysis of the existing defense called \fast{}~\cite{zhong2020neural} shows that using entity-based features capturing the factual structure of the text can lead to better adversarial robustness and generalization performance.
    
\end{itemize}

Our study provides many actionable insights that can be used 
to improve real-world applicability of defenses. 
Datasets and code used in this study are available on \textit{GitHub}.\footnote{\href {https://github.com/jmpu/DeepfakeTextDetection}{https://github.com/jmpu/DeepfakeTextDetection}}

\section{Background and Goals}\label{sec:background}
\noindent In this work, \textit{synthetic text} refers to text generated by DNN-based
LMs, and \textit{real text} refers to human-written text.


\para{Using language models for text generation.}
Synthetic text is generated using an LM. An LM is a statistical model that provides the joint probability distribution for a sequence of $n$ tokens $x_1, \cdots, x_n$. Tokens can be characters, words, or subword tokens~\cite{sennrich2015neural}. This joint distribution can be factorized by computing the conditional probability for each token, given the previous tokens: 
\begin{equation}\label{eqn: language_model}
    p(x_0, \cdots, x_n) = \prod\limits_{t=0}^n p(x_{t+1}~|~x_0, \cdots, x_t)
\end{equation}
Given a LM that provides the conditional probability in Equation~\ref{eqn: language_model}, synthetic text is generated using the following steps: Feed an initial \textit{priming sequence}, which can be a single token, $x_0$, into the LM, which then provides the conditional probability for the next token as $p(x_1~|x_0)$. This priming sequence can also be a special ``start-of-text'' token (which the model is trained to recognize) or a sequence of tokens. 
In the second step, the next likely token, $x_1$, is sampled from the distribution over the token vocabulary---a process known as \textit{decoding}. 
 Next, we can generate the next token, by feeding $(x_0, x_1)$ back into the LM, and sampling $x_2$, using the same decoding strategy. By repeating this process, one can generate text until a desired sequence length is reached, or an ``end-of-text'' token is chosen (which the model is trained to produce). 

Two key factors impact the quality of synthetic text---the decoding function, and the LM architecture: 

\para{Text decoding strategies.} Decoding strategies have witnessed significant development in recent years, and are now capable of producing high-quality synthetic text. Two effective decoding strategies are Top-k sampling~\cite{fan-etal-2018-hierarchical} and Top-p or Nucleus sampling~\cite{holtzman2019curious}. In Top-k sampling, the distribution is truncated and re-normalized to keep the $k$ most probable tokens, and then a token is randomly sampled. Holtzman et al.~\cite{holtzman2019curious} developed Top-p (or Nucleus) sampling, which produces more diverse and high quality text than Top-k sampling. In Top-p sampling, one truncates the distribution to keep the most probable tokens, such that their cumulative probabilities are greater than or equal to $p$, where $p$ $\in [0,1]$.
Another approach, Temperature sampling~\cite{ackley1985learning}, shapes the probability distribution by dividing the logits by a temperature parameter before passing them to the softmax function.
Low temperatures make the model produce less diverse text, as it makes more confident predictions (tokens), while temperatures greater than 1, result in more diverse text, as confidence is decreased.
The simplest decoding strategy is greedy sampling, where the most probable next token is chosen. However, temperature sampling and greedy sampling tend to produce repetitive text~\cite{holtzman2019curious}. 
Other sampling strategies such as beam search sampling
~\cite{wiseman2017challenges} are also known to suffer from similar problems~\cite{holtzman2019curious}.

\para{DNN-based LM architectures.} 
Until 2017, the de facto choices were RNNs~\cite{rumelhart1985learning} and LSTMs~\cite{graves2012long}. These models use a recurrent loop to maintain an internal ``hidden state'' that stores information about previous tokens, which is used to compute the next token distribution. However, they have limitations, including vanishing/exploding gradients~\cite{hochreiter1998vanishing}, the sequential nature of the model limiting parallelization, and an inability to generate longer coherent text~\cite{bengio1994learning}.

In 2017, Vaswani et al. proposed the Transformer~\cite{vaswani2017attention} architecture to address limitations of RNNs. Transformers, in contrast with RNNs/LSTMs, are based on the ``attention'' mechanism~\cite{bahdanau2014neural}. Instead of using a single hidden state to represent all previous tokens, attention mechanisms allow the model to compute vector representations of each token separately. These representations take into account context and relationship with all other tokens. Attention mechanisms can be customized to ``pay attention'' to only previous tokens (unidirectional attention), or to pay attention to future tokens, assuming they are provided (bidirectional attention). While the bidirectional attention is not suitable for text generation (future tokens are not available when generating text), it is useful when computing representations for other NLP tasks, \eg classification. Another advantage of Transformers is that each token can be processed in parallel, and is not dependent on previous tokens. This is made possible by using the teacher-forcing paradigm~\cite{Williams1989ALA}.
Transformers now power state-of-the-art for many NLP tasks. Examples include the popular \bert{}~\cite{devlin2019bert}, RoBERTa~\cite{liu2019roberta}, GPT-2~\cite{radford2019language}, and GPT-3~\cite{brown2020language} models.

\para{Goals.} Our goal is to understand the real-world applicability of existing defenses to detect synthetic text. We focus on two directions: (1) \textit{Understanding and improving performance of defenses in the wild.} The research community has made significant progress in developing detection schemes~\cite{zellers2019defending,ippolito2019human,gehrmann2019gltr,zhong2020neural,yao-2017-dnnreviews}. However, these defenses have been primarily tested on synthetic text produced by researchers themselves. It is unclear how well these methods would generalize to synthetic text in the wild, \ie those produced by the Internet community. We collect real-world synthetic text to understand the performance of existing detection schemes. We also propose efficient methods that adapt existing defenses to improve performance in the wild. (2) \textit{Understanding and improving performance against adaptive attackers.} Before deployment, defenders should consider an adaptive attacker who is knowledgeable about the defense and aims to evade detection. Existing works on evasion strategies primarily focus on generic black-box attacks that require a large number of queries to the defender's model~\cite{jin2020bert}. However, the defender's model may not even be exposed as a public API for queries. Also, one can detect query-based adversarial sample crafting schemes by looking for specific query patterns~\cite{chen2020stateful, li2020blacklight}. Using a surrogate model and relying on the adversarial samples to ``transfer'' may not always be effective either~\cite{yuan2021transferability, li2020bert}. Instead, we investigate more practical, (computationally) low-cost evasion strategies that require no queries to the defender's model, and no surrogate model as well. We also investigate methods to improve resilience against the proposed evasion strategies. 

\section{Models, Datasets and Metrics}
\label{sec: model_datasets_metrics}
\noindent We study 6 state-of-the-art defenses for detecting synthetic text.
To study synthetic text in the wild, we introduce 4 new synthetic datasets. We present metrics to evaluate defense performance on real-world datasets and against adaptive attackers.



\begin{table*}
    \centering
    \setlength{\tabcolsep}{4pt}
    \setlength\extrarowheight{3pt}
    \begin{tabular}{c|c|c|c|c|c|c|c|c|c}
    \hline
    \multirow{2}{*}{\textbf{Defense Models}} & \multicolumn{2}{c|}{\textbf{Train \& Test Datasets}} & \multirow{2}{*}{\textbf{\makecell{Decoding\\Setting}}} & \multirow{2}{*} {\textbf{\makecell{Train Set \\ Size/Class}}} & \multirow{2}{*}{\textbf{\makecell{Test Set \\ Size/Class}}} & \multicolumn{4}{c}{\bf{Performance (\%)}}\\
    \cline{2-3}
    \cline{7-10}
    & \bf{Real} & \bf{Synthetic} & &&& \bf{F1}& \bf{P}& \bf{R} & \bf{AUC} \\
    \hline
    \gbert{} & WebText~\cite{radford2019language} &GPT2-Large~\cite{radford2019language} & Top-p 0.96 & \makecell{10,000}& 4,000 & 88.8&91.7&86.0 &95.9\\  
    \hline
    \gltrgpt{} & WebText & GPT2-XL~\cite{radford2019language} &\makecell{Top-k 40\\ Temp 0.7} & 4,000 & 4,000& 98.5 & 98.9& 98.1 & 99.8\\
    \hline
    \gltrbert{} & WebText & GPT2-XL & \makecell{Top-k 40\\ Temp 0.7} & 4,000&4,000&79.6&78.7&80.6& 86.5\\ 
    \hline
    \grover{} & RealNews~\cite{zellers2019defending} &\grover{}~\cite{zellers2019defending}& Top-p 0.94  & 5,000 &4,000&87.1&83.4&91.1&94.3\\
    \hline
    \fast{} & RealNews & \grover{} & Top-p 0.96 & 5,000& 4,000&87.0&83.8&90.4&93.7\\
    \hline
    \rbert{} & RealNews & \grover{} & Top-p 0.96 & 5,000& 4,000&86.3&81.6&91.7&93.9\\
    \hline
    \end{tabular}
    \caption{Details regarding the training and evaluation of the defenses. From left to right: Datasets used for training and testing defense models; Decoding strategy used to generate synthetic data; Number of samples in training and test datasets; Detection performance of the defenses (F1, P, R, AUC represents F1 score, Precision, Recall, AUC ROC score, respectively).
    }\label{tab:defenses-setup}
\end{table*}

\subsection{Defenses: Synthetic Text Detection Schemes}\label{sec:six_defenses_info}
\noindent Existing defenses are supervised learning schemes that use a LM to extract features to build a binary classifier (real vs.~synthetic). We consider 5 existing defenses, including \grover{}~\cite{zellers2019defending}, \gltrbert{}~\cite{gehrmann2019gltr}, \gltrgpt{}~\cite{gehrmann2019gltr}, \gbert{}~\cite{ippolito2019human}, and \fast{}~\cite{zhong2020neural}. Inspired by \gbert{}, we build an additional defense, called \rbert{}. 
\para{Defense performance metrics.} Existing defenses are evaluated using a variety of metrics. \gbert{} reports accuracy and AUC, whereas \gltr{} reports only AUC. \fast{} and \grover{} use a modified version of accuracy, called paired accuracy (in addition to normal accuracy). Paired accuracy is computed by pairing real and synthetic articles such that both articles share the same metadata, \eg article title. If the detector assigns the synthetic article a higher probability than the real class, it is considered to be correctly classified. However, the paired accuracy setting is unrealistic because it assumes access to real articles used to generate the synthetic articles. Therefore, we do not use it in our study. We report all results primarily using the F1 score, Precision, and Recall for the synthetic class. These metrics provide insights into class-specific detection performance (unlike accuracy) and does not delegate the task of calibrating a decision threshold to future work. Additionally, we report AUC ROC scores for the 6 defenses on their original test datasets in Table~\ref{tab:defenses-setup}.

\para{Defenses.} Table~\ref{tab:defenses-setup} provides an overview of the training setup for each defense, and their performance.
\grover{}, \rbert{}, and \fast{} are trained to detect synthetic news articles, while \gltrbert{}, \gltrgpt{}, and \gbert{} are ``open-domain'' schemes and applicable to diverse topic domains.

\textit{(1) \grover{}.} Proposed by Zellers \etal{}~\cite{zellers2019defending}, it is a framework that can both generate and detect synthetic news articles. They first train a synthetic news article generator using a GPT-2 based LM. This generator is trained on the RealNews dataset~\cite{zellers2019defending}, a large corpus (120GB) of news articles from Common Crawl~\cite{common_crawl_link}. 
The GPT-2 model is modified to incorporate context fields for relevant metadata, \eg date, author names, article title, web domain and body text. This generator can produce synthetic news articles conditioned on the metadata. Next, to build a detection scheme, a classification layer
is attached to extract information from the hidden state of the special [CLS] token (placed at the end of each article). This updated model is fine-tuned on synthetic articles generated by \grover{} and more real articles from the RealNews dataset. \grover{} performs well when detecting text generated by \grover{} itself. However, this is not a realistic setting, as the defender is unlikely to have access to the attacker's generator. We study \grover{} in more realistic settings.


\textit{Our \grover{} setup:} We use the largest, publicly released version (1.5B parameters) of the \grover{} classifier called \grover{}-Mega~\cite{rowanzgr2:online}. Details of the training and testing setup are in Table~\ref{tab:defenses-setup}. \grover{} is fine-tuned using 5000 real articles from the RealNews dataset, and 5000 synthetic articles generated by \grover{} itself.
Using this publicly available model, we obtain an F1 score of 87.1\% on their original test set~\cite{zellers2019defending}.

\textit{(2) \gltrbert{} and \gltrgpt{}.} \gltr{}~\cite{gehrmann2019gltr} uses the insight that decoding strategies tend
to sample tokens that are assigned high probabilities by the LM. Hence, synthetic text can be detected by analyzing the likelihood of tokens in the text sequence, as determined by a LM. Presence of many high probability tokens is an indication that the text sample is likely synthetic. Using a LM, \gltr{} extracts features based on the number of tokens in the Top-10, Top-100, and Top-1000 ranks as determined by the token probability distributions. The features are then fed to a Logistic Regression classifier.


\textit{Our \gltr{} setup:} We use the authors' code to extract the features~\cite{gltr_github_link}, but no code for building the Logistic Regression classifier was released. The authors reported using Logistic Regression with default settings in the scikit-learn library~\cite{scikit-learn}. 
We additionally apply grid search on the hyperparameters to ensure the classifier is properly tuned (See Appendix~\ref{sec:gltr_grid_search} for details).
To build an open-domain classifier, we train on synthetic text from GPT2-XL (similar to the original work) and real articles from the WebText dataset~\cite{radford2019language}. Both sources are known to cover diverse topics.
Similar to the original work~\cite{gehrmann2019gltr}, we create 2 variants of \gltr{}, namely \gltrbert{} that uses \bert{}~\cite{devlin2019bert}, and \gltrgpt{} that uses the GPT2-XL~\cite{radford2019language} as the back-end LM.\footnote{\gltr{}~\cite{gehrmann2019gltr} used GPT2-small as the LM, but we use a larger and more accurate LM, GPT2-XL.} 
We obtain F1-scores of 98.5\% and 79.6\% for \gltrgpt{} and \gltrbert{}, respectively.

\textit{(3) \gbert{}.} A \bert{}-based binary classifier, proposed by Ippolito \etal{}~\cite{ippolito2019human}, attaches a classification layer to a pre-trained \bert{}-Large LM, and then fine-tunes it on a dataset of synthetic and real articles. 

\textit{Our \gbert{} setup:} The authors did not release the datasets and models. We replicated their experimental setup.
See Table~\ref{tab:defenses-setup} for details. 
While the original work reported an accuracy of 81\%, our model achieves an F1 score of 88.8\% on the test set. We achieve a higher F1 score, even after using a smaller training set (10,000 articles per class versus 250,000 articles in the original work). Our model uses the full context window size of 512 tokens, supported by BERT, unlike the smaller window size of 192 tokens in the original. We suspect the higher performance is likely due to the larger context window used by our implementation.

\textit{(4) \rbert{}.} Inspired by \gbert{}, we create an additional defense using the same approach, but with a different language model, RoBERTa~\cite{liu2019roberta}. RoBERTa makes several changes to the \bert{} LM, such as training the model on a larger dataset with a bigger batch size, removing the next-sentence-prediction task, training on longer sequences and dynamically changing the masking pattern applied to the training data. RoBERTa is known to outperform \bert{} on NLP tasks such as GLUE~\cite{wang2018glue}, SQuAD~\cite{rajpurkar2016squad}, and RACE~\cite{lai2017race}.


\textit{Our \rbert{} setup:}
We train a RoBERTa-base model on synthetic text produced by \grover{} and real text obtained from the RealNews dataset, and obtain an F1 score of 86.3\% in detecting synthetic news articles. 





\textit{(5) \fast{}.} \fast{}, proposed by Zhong \etal{}~\cite{zhong2020neural}, unlike other defenses, taps into the semantic layer or the ``factual structure'' of text. 
\fast{} uses a graph-based learning approach that uses features based on named entities in the document.
\fast{} exploits the insight that state-of-the-art text generators still produce inconsistencies in the factual structure of text, when compared to real text. For example, while it is easy for humans to correctly mention and reference named entities (\eg location, people, objects) across sentences, a text generator might make mistakes and create inconsistencies in how entities are mentioned in continuous sentences. To capture such inconsistencies, \fast{} constructs an entity network based on how entities are referenced within and across sentences, and uses a graph convolution network (GCN) to learn patterns in the network. In addition, \fast{} also uses the RoBERTa LM~\cite{liu2019roberta} to extract token, and document-level representations, in conjunction with the GCN-based features. We analyze \fast{} in detail in Section~\ref{DistilFAST}.


\textit{Our \fast{} setup:} A pre-trained model for \fast{} was not available. We obtained code for \fast{}, and reproduced the experimental setup described in the original work. Similar to the original work, we train \fast{} to detect synthetic news articles, and train it on the RealNews dataset (real class) and text from \grover{} (synthetic class).~\footnote{Zhong \etal{}~\cite{zhong2020neural} additionally trained a version of \fast{} on WebText (real) and GPT2-XL text (synthetic). We omitted this setting for simplicity.} We obtain an F1 score of 87\% in detecting news articles (Table~\ref{tab:defenses-setup}).

\para{Context window size for training and evaluation.} 
All 6 defenses use a context window size of the first 512 tokens in an article
for detection, as determined by its tokenization scheme.
To understand the impact of the context window size for detection, we evaluate all defenses against smaller context window sizes, \ie 64, 128, 256 tokens, and present results in Figure~\ref{fig:baseline_token_length} in the Appendix.
Performance (F1 score) monotonically increases as the context window size increases. Therefore, we chose a 512 token window size. Larger window size would significantly increase computational complexity for all experiments, and some pre-trained models only support a certain maximum window size (\eg 512 tokens for BERT).


\para{Other defenses.} 
There are a few other defenses that are not considered in our study. For example, Yao \etal{}~\cite{yao-2017-dnnreviews} proposed a method to detect LSTM-generated synthetic reviews. We omit it because our preliminary evaluation yielded unsatisfactory results---an F1 score of only 68\% in detecting GROVER generated text. We discuss more details of these experiments and other defenses in the Appendix~\ref{sec:other_defenses}.

\subsection{Evaluation Metrics}
\label{adaptive_attack_eval_metric}
\noindent We use the following metrics to measure defense performance on text in the wild and against adaptive attackers:

\para{Percentage change in detection performance: $\Delta$F1, and $\Delta$Recall.} We measure the percentage change in detection performance for the synthetic class, when applied to a new test dataset, compared to a specific baseline performance. This is broken into percentage changes in F1 and Recall, \eg $\Delta F1=(F1_{new}-F1_{baseline})/(F1_{baseline})$. The new test dataset can be an \textit{In-the-wild} dataset, or a dataset containing synthetic text produced by an adaptive attacker. We define baseline performance depending on the experiment context, which usually refers to the defense performance when evaluated on the test datasets considered in the original work (\eg numbers in Table~\ref{tab:defenses-setup}).
For attack experiments, where only synthetic samples are modified in the test set (\ie real samples are the same in the test and baseline settings), we only consider $\Delta$ Recall (for the synthetic class), as there will be no change in false positives (real samples classified as synthetic). Note that the change in performance can be a degradation or improvement in performance.

\para{Evasion rate (ER).} In some attacks, the adaptive attacker perturbs existing synthetic samples to evade detection. In such settings, we use evasion rate, defined as the fraction of perturbed synthetic samples that evade detection by a defense. Higher fraction indicates higher attack success.

\para{Evaluating quality of synthetic text.} For adaptive attackers, it is not sufficient to evade detection, it is also necessary to maintain high linguistic quality of the generated text. We measure linguistic quality using the state-of-the-art GRUEN metric~\cite{zhu2020gruen}. Zhu \etal{} proposed GRUEN, an unsupervised, reference-less metric designed for synthetic text. GRUEN correlates highly with human judgements, and better than other existing metrics for linguistic quality. The metric, computed for a synthetic sample, ranges from 0 to 1, and a higher value indicates better linguistic quality. An advantage is that this metric does not require any reference text, which usually requires human effort to obtain. GRUEN measures linguistic quality based on ``grammaticality'', non-redundancy, discourse focus, structure and coherence.\footnote{The GRUEN implementation we obtained from the authors, does not compute structure and coherence. The authors claim that this omission does not impact the metric scores significantly.} More details are in the Appendix~\ref{gruen_subscores}.
Other linguistic quality metrics have also been proposed over the years.
In the Appendix~\ref{other_text_quality_metrics} we describe other metrics,
and justify our choice of using GRUEN over the other metrics.

\begin{table}
    \centering
    \setlength\extrarowheight{3pt}
    \begin{tabular}{c|c|c}
    \hline
    \multirow{2}{*}{\bf Datasets} &  
    \multirow{2}{*}{\bf \makecell{\#Documents\\per Class}} &
    \multirow{2}{*}{\bf Document Topic(s)}\\
    &  & \\
    \hline
    AI-Writer & 1,000 & News\\
    \hline
    ArticleForge & 1,000 & News\\
    \hline
    Kafkai & 1,000 & \makecell{Cyber Security,\\ SEO, Marketing}\\
    \hline
    RedditBot & 887 & Reddit Comments\\
    \hline
    \end{tabular}
    \caption{Details of the \textit{In-the-wild} datasets.}
    \label{tab:in_the_wild_details}
\end{table}

\subsection{In-the-wild Datasets}
\label{sec:in_the_wild_datasets}
\noindent 
We collect 4 \textit{In-the-wild} datasets from the web containing both synthetic and real articles from matching semantic categories. All measurements were conducted from Nov. 2020 to Apr. 2021. This includes synthetic text posted by Internet users, and text from synthetic \textit{text-generation-as-a-service} platforms, geared towards the SEO community. Synthetic text generation services could be misused to create fake news articles, fake reviews, or fake web articles for BlackHat SEO activities. While we could not verify the text generators used by the services we study, they claim to use customized versions of Transformer-based LMs.
This again highlights the need to understand real-world performance of defenses, because text generators used in the wild can be different from those used by the research community. 
Table~\ref{tab:in_the_wild_details} shows dataset statistics.

\para{AI-Writer.} This dataset was collected from the text-generation service, AI-Writer~\cite{ai_writer}. 
Given a title, AI-Writer claims to generate factually accurate articles capturing recent information on the topic (based on the title). 
In our email communication with the service, they claim to employ custom Transformer-based LMs that are not available off-the-shelf. 
Since AI-Writer requires titles to generate articles, we first collect real news articles, and use the title from the real articles for generation. We evenly and randomly scraped 1000 real news articles from 20 popular news websites sampled from the RealNews~\cite{zellers2019defending} dataset.
The list of websites is shown in Table~\ref{tab:ai_writer_aforge_websites_appendix} in the Appendix. We verified that this dataset has no overlap with the training datasets of the defenses (Table~\ref{tab:defenses-setup}), based on the article publication dates.
These articles form the real class of the dataset. Next, we used the real article titles to generate 1000 synthetic articles from AI-Writer. 
AI-Writer charges for article generation, and we spent \$400 for generating 1000 articles. 
\para{ArticleForge.} This dataset was collected from the ArticleForge text-generation service~\cite{articleforge}. ArticleForge requires a set of keywords to generate an article. 
As per our communication with the service, they claim to use fine-tuned versions of GPT-2~\cite{radford2019language}, \bert{}~\cite{devlin2019bert} and T5~\cite{raffel2020exploring} to generate synthetic text. We follow a similar methodology as used for AI-Writer, and collect 1000 synthetic, and 1000 real news articles. 
ArticleForge charged us \$57, for which they allow unlimited article generation for a month. 
\para{Kafkai.} This dataset was collected from the Kafkai text-generation service~\cite{kafkai}. Given one of 25 categories and an initial priming text, Kafkai generates a unique synthetic article that belongs to that category and is contextualized by the priming text. As per our communications with their service, Kafkai uses models from OpenAI, including GPT-2, and fine-tunes them on millions of articles to generate high quality synthetic text. We follow a similar methodology as AI-Writer and ArticleForge, and obtain context for the synthetic article generation from 1000 real articles---100 articles from 10 of the 25 available categories, \eg Cybersecurity, SEO, and Marketing. 
We use the first 50-100 words from each of the 1000 real articles as priming text to generate 1000 articles from Kafkai. Priming text is not included within the final article.
Kafkai charged us \$129 for generating 1000 articles.




\para{RedditBot.} This dataset was collected from Reddit.com. A GPT-3 powered bot posted comments under the username /u/thegentlemetre and interacted with 
users on /r/AskReddit, a popular subreddit on Reddit.com. Real Reddit users were initially unaware that it is a bot, and resulted in interactions with users for a week~\cite{gpt3_redditbot}. 
We collected 1,204 comments posted by the bot between 27th Sep, 2020 and 15th Apr, 2021, and retained 887 comments with a length greater than 192 tokens (we discard synthetic comments that are too short in interactions). 
We then scraped 112,296 real comments in every forum thread that contained a bot comment. To create a balanced dataset, we use a random sample of 887 real comments (with at least 192 tokens).

\para{Linguistic quality of synthetic text in the wild}
We use the GRUEN metric to evaluate linguistic quality. We observe that synthetic data in the wild is comparable to synthetic text produced by the research community. More details of the text quality comparison are presented in the Appendix~\ref{sec:wild_text_quality}. 

\section{Defense Performance in the Wild}\label{Perf_In_wild}

\begin{table*}
    \centering
    \setlength{\tabcolsep}{1pt}
    \setlength\extrarowheight{3pt}
    \begin{tabular}{P{1.8cm}|P{1.2cm}|P{1.2cm}|P{1.2cm}|P{1.2cm}|P{1.2cm}|P{1.2cm}|P{1.2cm}|P{1.2cm}|P{1.2cm}|P{1.2cm}|P{1.2cm}|P{1.2cm}}
    \hline
         \multirow{2}{*}{\bf Datasets} & \multicolumn{2}{c|}{\bf \gbert{}} & \multicolumn{2}{c|}{\bf \gltrgpt{}} & \multicolumn{2}{c|}{\bf \gltrbert{}} & \multicolumn{2}{c|}{\bf \grover{}} & \multicolumn{2}{c|}{\bf \fast{}}& \multicolumn{2}{c}{\bf \rbert{}}  \\
         \cline{2-13}
         & \bf F1 &\bf  $\Delta$F1 & \bf F1 &\bf  $\Delta$F1 &\bf  F1 &\bf  $\Delta$F1 &\bf  F1 &\bf  $\Delta$F1 & \bf F1 &\bf  $\Delta$F1 & \bf F1 &\bf  $\Delta$F1 \\
         \cline{2-13}
         \hline
         AI-Writer & 28.8 & -67.6 & 1.6 &-98.4 & 64.8 & -18.6 & 87.6 & +0.6& 94.9 & +9.1 & 92.1 & +6.7 \\
         ArticleForge & 19.7 & -77.8 & 44.1& -55.2& 85.6 & +7.5 & 76.9& -11.7& 88.5 & +1.7 & 87.4 & +1.3 \\
         Kafkai & 65.9 & -25.8 &1.0 &-99.0 & 41.4 & -48.0 & -- & -- & --&--&--&--\\
         RedditBot & 14.1 &-84.1  & 61.5& -37.6& 83.4 & +4.8 & -- & -- &--&--&--&--\\
         \hline
    \end{tabular}
    \caption{Performance of the defenses, \ie F1 (\%) and $\Delta$F1 (\%) of the synthetic class, on the \textit{In-the-wild} datasets. 
    The percentage change in F1 ($\Delta$F1) is computed from the baseline performance of each defense. `+' means performance improvement and `-' means performance degradation. ``--'' (the longer minus mark) indicates experiments we ignored: We did not test \grover{}, \fast{} and \rbert{} on non-news domain datasets, \ie Kafkai and RedditBot. This is because these defenses are only trained for the news domain.
    }
    \label{tab:eval_in_the_wild}
\end{table*}

\subsection{Detection Performance}
\label{sec:detection_performance_defenses}
\noindent We test the 6 defenses (Section~\ref{sec:six_defenses_info}) on our 4 \textit{In-the-wild} datasets (Section~\ref{sec:in_the_wild_datasets}). Since \grover{}, \fast{} and \rbert{} are trained for the news domain, we only test them on news domain datasets, which includes AI-Writer and ArticleForge. The remaining defenses, \gltrbert{}, \gltrgpt{}, and \gbert{} are trained on a diverse corpus and therefore can be tested on all the datasets.
We report F1 and $\Delta$F1 (Section~\ref{adaptive_attack_eval_metric}). To compute $\Delta$F1, we use the performance of each defense on their original test set as the baseline performance (see Table~\ref{tab:defenses-setup}). 

Detection performance in the wild is presented in Table~\ref{tab:eval_in_the_wild}. Detailed results including the Precision and Recall scores are in the Appendix (Table~\ref{tab:eval_in_the_wild_long}). Before we discuss the results, note that all 6 defenses achieve high detection performance (79.6\% to 98.5\% F1) on their original test datasets (Section~\ref{sec:six_defenses_info}). Our key findings are as follows:

\para{\textit{Finding 1:}} 
\textit{Open-domain defenses show significant degradation in performance when applied to synthetic text in the wild, while defenses trained on data from a specific domain are able to detect In-the-wild data from that domain.} All three open-domain defenses --- \gbert{}, \gltrbert{}, \gltrgpt{} show significant performance degradation ranging from 18.6\% to 99.0\% degradation in F1 score. All 3 of these defenses exhibit performance worse than a random predictor (50\% F1) for at least one \textit{In-the-wild} dataset. \gbert{} and \gltrgpt{} show significant degradation on all the datasets. All three news-based defenses --- \grover{}{}, \rbert{}{}, \fast{}{} perform well above the open-domain defenses on the news-based datasets. \fast{} and \rbert{} perform better on these datasets than on their original test datasets.

We further investigate the performance differences between the open-domain and the news domain defenses. Our hypothesis is that this can be attributed to distribution shift or distributional differences between data in the wild and the original datasets used to train/evaluate the defenses. To study this, we choose \gbert{} from the open-domain category, and \grover{} from the news category. We use a simple metric, average-linkage~\cite{moseley2017approximation}, to measure the distribution distance between two synthetic datasets. Given two synthetic datasets, $X$, and $Y$, we define the distribution distance as $D(X,Y)$. To represent a dataset's distribution, we randomly select 1,000 articles (or all available samples if there are fewer samples) from the dataset, and then extract the hidden state of the special [CLS] token (used as the input to the classifier) in each article from the detector as its representation. Thus $X$ and $Y$ each includes 1000 embedding vectors. Larger values of $D(X, Y)$ indicate a larger distribution shift.

\begin{figure}[!t]
\centering
\includegraphics[width=0.95\columnwidth]{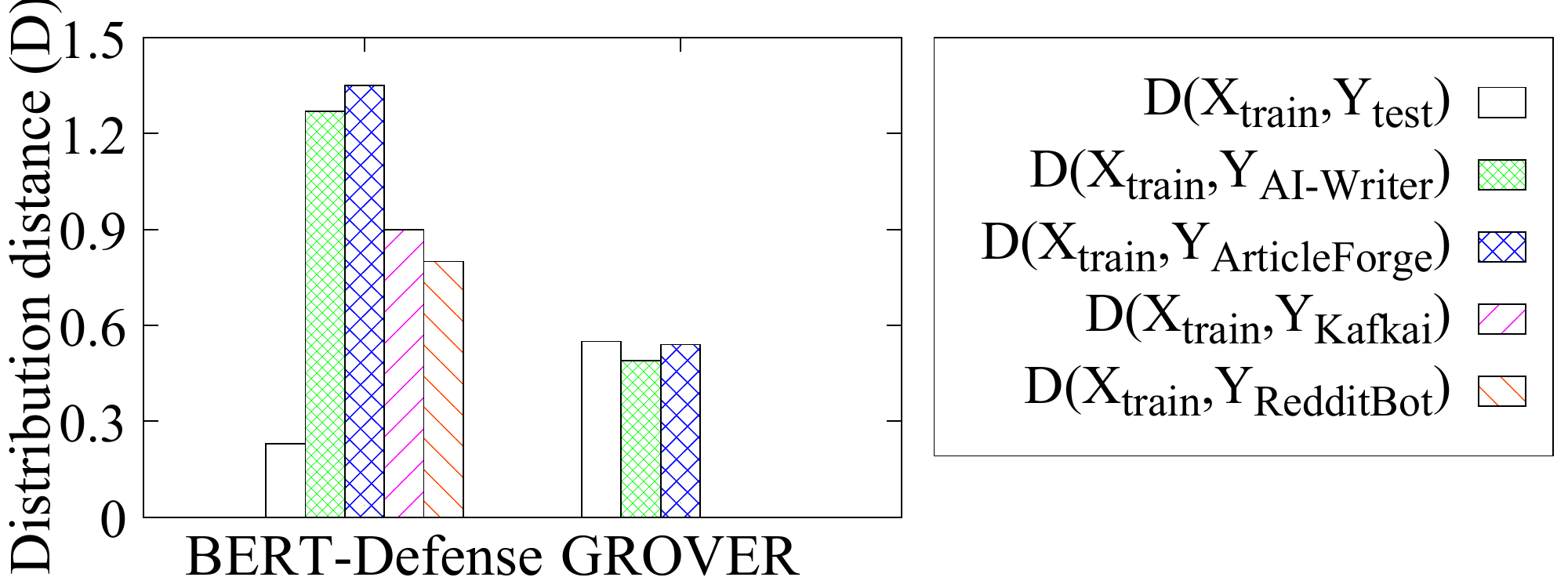}
\caption{Distribution distance between the training set and the test sets (including baseline test set and \textit{In-the-wild} test sets) of \gbert{} and \grover{}.}
\label{fig:dist_bar}
\end{figure}

For each defense, we compute 2 types of distance measures: (1) Distance $D(X_{train}, Y_{test})$ between its training dataset $X_{train}$ and its original test set $Y_{test}$ (\ie test set used in Table~\ref{tab:defenses-setup}). (2) Distance $D(X_{train}, Y_{wild})$ between its training dataset and each of the In-the-wild datasets $Y_{wild}$.
We expect that $D(X_{train}, Y_{wild})$ is greater than $D(X_{train}, Y_{test})$ if the defense's performances on \textit{In-the-wild} datasets degrade, and $D(X_{train}, Y_{wild})$ is closer to $D(X_{train},Y_{test})$ if the detection performances are similar. 
Results are in Figure~\ref{fig:dist_bar}. All the $D(X_{train},Y_{wild})$ are significantly greater than $D(X_{train},Y_{test})$ for \gbert{}, but not for \grover{}. This observation is in line with our hypothesis.


\para{\textit{Finding 2:} } \textit{Robustly pre-trained bidirectional models, \ie RoBERTa, generalize better than unidirectional models.} \rbert{} and \fast{} show performance improvements (F1) when applied to synthetic news samples in the wild. Both approaches use features extracted using RoBERTa, which improves over the BERT bidirectional model. The authors of RoBERTa show that BERT is significantly undertrained, and propose changes to improve BERT's pre-training~\cite{liu2019roberta}. It is worth noting that \grover{}, a unidirectional model, claims to perform better than bidirectional models (\eg BERT). However, our finding suggests that this claim is not true if the bidirectional model is robustly pre-trained, as in the case of RoBERTa. Moreover, a unidirectional model like \grover{} is over 10x larger than the RoBERTa-base we use, in terms of number of parameters, yet it under-performs.

Another surprising result is that of \gltrbert{} performing well on ArticleForge and RedditBot (a GPT-3 dataset), but \gltrgpt{} performs poorly on these two datasets. This means that the back-end LM model (BERT vs GPT-2) used by GLTR can have a huge impact on generalization performance. 

\begin{table*}
    \centering
    \setlength{\tabcolsep}{6pt}
    \setlength\extrarowheight{4pt}
    \begin{tabular}{c|c|c|c|c|c|c|c|c}
    \hline
         \multirow{4}{*}{\bf Datasets} &
         \multicolumn{8}{c}{\bf Detection Performance (F1)}\\
         \cline{2-9}
         &\multicolumn{4}{c|}{\bf \gbert{}} & \multicolumn{4}{c}{\bf \grover{}} \\
         \cline{2-9}
         & \multirow{2}{*}{\bf \makecell{Before \\ Fine-tuning}} & \multicolumn{3}{c|}{\bf \makecell{\# Samples\\for Fine-tuning}} & 
         \multirow{2}{*}{\bf \makecell{Before \\ Fine-tuning}} &
         \multicolumn{3}{c}{\bf \makecell{\# Samples\\for Fine-tuning}}\\
          \cline{3-5}
          \cline{7-9}
         &&\bf 10&\bf 50&\bf 100&&\bf 10&\bf 50&\bf 100\\
         \cline{2-9}
         \hline
         AI-Writer & 28.8&68.9&85.8&90.4&87.6&90.7&93.8&94.8\\
         ArticleForge &19.7&85.3&90.8&93.7&76.9&86.5&95.9&97.1\\
         Kafkai &65.9&71.0&83.8&85.0&--&--&--&--\\
         RedditBot &14.1&71.5&90.6&95.2&--&--&--&--\\
         \hline
    \end{tabular}
    \caption{Detection performance in F1 (\%) of \gbert{} and \grover{} on \textit{In-the-wild} datasets before and after fine-tuning on a limited set of articles from the real and synthetic classes. ``--'' represents experiments we ignored: \grover{} is a news domain defense, and not applicable to non-news domain datasets, \ie Kafkai and RedditBot.
    }
    \label{tab:fine_tune_wild}
\end{table*}

\subsection{Improving Performance in the Wild}
\label{sec:improve-in-the-wild}
\noindent Can we adapt the defenses that currently perform poorly (in the wild) to perform better on a target dataset? We investigate domain adaptation via transfer learning, \ie by fine-tuning the classifier on data from the target distribution. Language model fine-tuning has shown tremendous success for domain adaptation~\cite{howard2018universal}. We consider a realistic and challenging setting, where the samples from the target distribution (for fine-tuning) are limited---in our case, as little as 10, 50 or 100 samples each for the synthetic and real class. This fits a scenario where the Internet community, including text-generation-as-a-service platforms, rapidly updates their generative models, or produces many model variants over time. In such a setting, it is hard to obtain abundant ground-truth data for attack class (synthetic samples). In such a setting, can the defender keep up?
 

We fine-tune the models by extending the training on the binary classification task. We consider the defenses, \gbert{} and \grover{} since they exhibit a degradation in performance (Table~\ref{tab:eval_in_the_wild}).\footnote{\gltr{} also exhibits a degradation in performance, but we do not consider it. \gltr{} is not a DNN-based classifier, and is therefore not suitable for fine-tuning.} 
While fine-tuning \gbert{} on our small datasets, we encountered a known issue of training instability~\cite{devlin2019bert}.
To overcome this, we employ the revitalization strategy proposed by Zhang et al.~\cite{zhang2020revisiting}. The training hyperparameters used in \gbert{} and \grover{} fine-tuning experiments can be found in the Appendix~\ref{sec:exp_details}. Results are in Table~\ref{tab:fine_tune_wild}.
Our findings are as follows:

\para{\textit{Finding 3:}} \textit{Fine-tuning with as limited as 10 In-the-wild data samples can help defenses adapt to new domains, and more samples lead to better fine-tuning performance.} 
Both defenses benefit from observing a few samples from both classes of the target dataset. Moreover, detection performance only improves with more fine-tuning samples. 

\section{Defense Against Adaptive Attackers}
\label{sec:adaptive-attacks}

\subsection{Attack Methods}
\label{sec:adaptive-attack-methods}

\noindent To ensure real-world applicability, defenses should be effective against adaptive adversaries who are aware of the defense scheme, and can adapt the synthetic text to bypass detection. We focus on low-cost and practical adaptive attacks. Our attacks do not require a computationally expensive re-training of the attacker's generative model or creation of a surrogate/shadow defense model to craft adversarial samples. We assume a black-box setting requiring no queries to the defense scheme to craft adversarial samples. 
We also consider maintaining the linguistic quality of the synthetic text as the attacker's constraint.
If linguistic quality is degraded significantly, it impacts the attacker's goals of misleading users, \eg synthetic fake news articles with poor linguistic quality could raise suspicion from users.
These assumptions provide a more realistic setting for attackers. Our attack methods are split into two categories: (1) Attacks that change the text generation process without re-training the text generator, and (2) attacks that add adversarial perturbations to existing synthetic text samples to evade detection.
    


\subsubsection{Evasion by Changing the Text Generation Process} 
\noindent Existing defenses are trained on synthetic text, created based on a specific decoding strategy and priming process. Our goal is to understand the robustness of defenses against changes to the text distribution triggered by varying the text generation process.
 Given a generative model, our idea is to craft different distributions of synthetic text samples by: (1) varying the text decoding method (and its parameters), and (2) by varying the number of priming tokens used by the model. 
 We evaluate the linguistic quality of the adapted synthetic text using the GRUEN metric (Section~\ref{adaptive_attack_eval_metric}). \textit{An attack is considered successful only if it degrades defense performance, while preserving linguistic quality or with limited degradation in linguistic quality.}

\begin{table*}
    \centering
    \setlength{\tabcolsep}{6.5pt}
    \setlength\extrarowheight{1.5pt}
    \begin{tabular}{c|c|c|c|c|c|c|c|c|c|c}
    \hline
         \bf \multirow{3}{*}{\bf Defenses} 
         & \bf \multirow{3}{*}{\bf{\makecell{Baseline \\ Decoding \\ Setting}}} 
         & \multicolumn{9}{c}{\bf Attack: Changing the Decoding Strategy}\\
         \cline{3-11}
         & & \multicolumn{3}{c|}{\bf Top-p Decoding} & \multicolumn{3}{c|}{\bf Top-k Decoding} & \multicolumn{3}{c}{\bf Temperature Decoding}\\
         \cline{3-11}
         & &\bf{Top-p} & \bf $\Delta$R & \bf $\Delta$\gruen{}  & \bf{Top-k} & \bf $\Delta$R & \bf $\Delta$\gruen{} & \bf{Temp} & \bf $\Delta$R & \bf $\Delta$\gruen{} \\
         \cline{3-11}
         \hline
         \gbert{}&Top-p 0.96 &0.8 & -13.3& +0.5& 40 & -12.5& +4.0&--&--&--\\
         \hline
         \gltrgpt{}&\makecell{Top-k 40\\Temp 0.7} &0.98 & -97.6& +2.0& 160 &-56.4 &+8.8 & 0.9 &-90.9 &+4.0 \\
         \hline
         \gltrbert{}&\makecell{Top-k 40\\Temp 0.7} & 0.98& -90.0& +2.0& 160& -50.7&+8.8&0.9 &-57.3& +4.0\\
         \hline
         \grover{}&Top-p 0.94 & 1.0&-35.6 & -4.1& 160 & -6.8&-0.5 &--&--&--\\
         \hline
         \fast{}&Top-p 0.96 &1.0& -9.7 &	-3.6& 80&  -2.9	& +0.0 &--&--&--\\
         \hline
         \rbert{}&Top-p 0.96 &1.0 &-22.0	&-3.6	& 160&	-2.4	&	+0.0   & -- & -- & --\\
         \hline
    \end{tabular}
    \caption{Performance of attacks that change the decoding strategy of the LM. Evaluation metrics include Recall (R) of the synthetic class and the average GRUEN (GRN) of synthetic data. We show the percentage change of each evaluation metric ($\Delta$R, $\Delta$GRN) from the baseline performance of each defense on the most effective attack of each decoding strategy. `+' means performance improvement and  `-' means performance degradation. ``--'' (the longer minus mark) indicates experiments we ignored: We only consider Temperature decoding for GLTR defenses, because the original GLTR work was evaluated using temperature-based decoding. 
    }
    \label{tab:decoding_eval_short}
\end{table*}

Each defense was trained to detect text produced using a certain decoding and priming strategy. We consider this as the baseline settings, shown in the second column of Tables~\ref{tab:decoding_eval_short} and~\ref{tab:priming_eval_short}. We vary the decoding strategy by considering the following methods: Top-k, Top-p (or nucleus sampling), and Temperature decoding. For each decoding strategy, we also consider different parameter settings. For Top-p, we consider several values of $p$ in the range $[0.8, 1]$ in small increments. For Top-k, we vary $k$ using the following values $[40, 80, 120, 160]$, and Temperature in the range $[0.7, 0.9]$. These values were determined based on standard value ranges used in prior work~\cite{zellers2019defending, zhong2020neural}, and 
values beyond this range resulted in significant degradation in linguistic quality.


All the defenses, except \gbert{} is trained on unconditional text (\ie number of priming tokens is 0). \gbert{} uses a single priming token. Priming language models with some tokens, as opposed to unconditional generation, can change the statistical and qualitative properties of generated text as the model might never generate the priming sequence on its own, \eg due to its low probability. For this attack, we generate text using a varying number of priming tokens $n$, where $n \in [1, 4, 8, 12]$. Each synthetic article is generated using the first $n$ tokens from a real article. We limit the maximum number of priming tokens to 12 to minimize the amount of real text in synthetic articles. 
\subsubsection{Evasion by Adversarial Perturbations}
\label{sec:dftfooler_tech_details}
\noindent We craft adversarial inputs in a black-box setting by leveraging insights unique to the synthetic text detection problem. \\
\para{Crafting adversarial inputs using \ourattack{}.}
Given a synthetic sample, our approach called \ourattack{}, aims to misclassify it as real by adding adversarial perturbations to it.
Unlike existing work on adversarial inputs in the text domain~\cite{morris2020textattack}, \ourattack{} requires no queries to the victim model, or a surrogate/shadow classifier to craft the perturbations. \ourattack{} only requires a pre-trained LM, and several versions are publicly available today~\cite{hugging_face_link}.

Given a synthetic article, we identify a (limited) set of words that are important for classification, and replace them with words that alter the model's prediction while preserving semantic similarity and linguistic quality. \textit{The challenge is in identifying the important words and finding suitable replacements that alters the prediction}---we draw insights from the \gltr{} approach. Our insight is that generative models tend to generate the next token from the head of the distribution, 
thus capturing only a limited subset of the true distribution of natural language~\cite{gehrmann2019gltr}. 
In other words, if we pass a synthetic article and a real article through a pre-trained LM (\eg BERT), the synthetic article is likely to contain many tokens which have a high probability of being generated by that language model. On the other hand, real articles will contain many low probability tokens since humans exhibit greater variation in their choice of words and this is hard for LMs to emulate. 
Our hypothesis is that defense schemes learn this difference between real and synthetic articles for discrimination. Therefore, to misclassify a synthetic sample, 
we replace a subset of the most confidently predicted words (using a pre-trained LM) by its synonyms that have lower confidence (according to the same LM).

First, \ourattack{} scans a given article to choose the top $N$, most confidently predicted words according to a chosen LM, for replacement. 
The importance of a word is determined by the absolute rank of its token probability predicted by the LM.~\footnote{For words that are split into multiple sub-tokens, we use the probability prediction of the first sub-token in the word.} \ourattack{} does not perturb stop words. More details about using a LM to find important words are described in the Appendix~\ref{sec:dft_details}.


The second step is to find replacement words, while preserving semantics. To find a synonym for replacement, we build on the methodology used by TextFooler~\cite{jin2020bert}, and adapt it to our setting to not require queries to the victim model. 
For a targeted word, there are 4 sub-steps to find a valid synonym: \textit{(1) Synonym Extraction:} We first extract a candidate set of synonyms for the targeted word as its possible word replacements. Synonym candidates are chosen according to the cosine similarity between the word embeddings of the targeted word and every other word in the vocabulary. We use word embeddings from~\cite{mrkvsic2016counter} which are specially curated for finding synonyms. \textit{(2) POS Checking}: The goal of POS checking is to assure that the grammar of the perturbed text remains the same. We will only keep synonyms with the same part-of-speech (POS) tag as the targeted word.
\textit{(3) Semantic Similarity Checking:} This step ensures that the sentence semantics before and after replacing the targeted word remain similar.
Similar to TextFooler, we use the Universal Sentence Encoder (USE)~\cite{cer2018universal} to compute sentence similarity.
At this step, we only keep synonyms that can maintain sentence similarity scores above 0.7.
\textit{(4) Choose a synonym with low confidence as measured by a LM:} At the last step, we choose a replacement word from valid synonyms that is predicted by the LM with a low probability of $\leq 0.01$. We choose the synonym with the lowest probability if multiple synonyms meet the threshold requirement.
It is possible that less than $N$ words are replaced if the semantic similarity conditions are not met. Empirically, we find that $N=10$ works well in practice, offering a trade-off between evasion rate, and preserving linguistic quality/semantics. We implement \ourattack{} using 2 back-end LMs, namely BERT and GPT2-XL. That said, the backend model is replaceable when more advanced LMs emerge in the future.

\para{Perturbation attack baselines.} We compare attack performance of \ourattack{} with the following two approaches: (1) \textit{TextFooler~\cite{jin2020bert}}: TextFooler is a black-box attack that requires a large number of queries to the defense model to craft adversarial perturbations. TextFooler is highly effective against Transformer-based classifiers, and also preserves the utility of the attack by preserving the semantic content. TextFooler finds important words to replace by querying the defense model, and also queries the model to find the best replacement words. (2) \textit{random perturbations:} This is a simple baseline that replaces random words in the article by synonyms that preserve the semantic content. This approach requires no LMs or queries to the defense model. Similar to \ourattack{}, we only replace $N$ words in an article. Such a baseline would serve to understand the benefit of replacing words based on their importance (as in our \ourattack{}). \ourattack{} should perform better than this baseline to be considered a useful attack.

\subsection{Attack Evaluation}
\label{sec:adaptive-attack-eval}
\noindent In this section, we evaluate the adaptive attacks.

\para{Evasion by adapting the decoding process}
Table~\ref{tab:decoding_eval_short} presents the results for adapting the decoding strategies. Attack success is measured using the percentage change in Recall for the synthetic class, $\Delta$R. To compute $\Delta$R, the baseline setting is the performance on the original test dataset of each defense (Table~\ref{tab:defenses-setup}). A higher $\Delta$R indicates better attack success.  We only report $\Delta$R because the real set is the same as in the baseline experiment setting (from Table~\ref{tab:defenses-setup}).
In Table~\ref{tab:decoding_eval_short}, for each defense, we show results for the most effective attack configuration (\ie decoding method and its parameters). The most effective attack is the one with the largest degradation in $\Delta$R, with only a minor (up to 5\%) or no degradation in linguistic quality measured by the GRUEN score. We report percentage change in average GRUEN score ($\Delta$GRN) of articles before and after changing the decoding strategy. For Temperature decoding, we only show results for the \gltr{} defenses, because \gltr{}  was originally evaluated using Temperature decoding. Temperature decoding is known to produce repetitive text~\cite{ackley1985learning}, and is omitted for the other defenses. Findings are as follows:

\begin{figure}[!t]
\centering
\includegraphics[width=0.65\columnwidth]{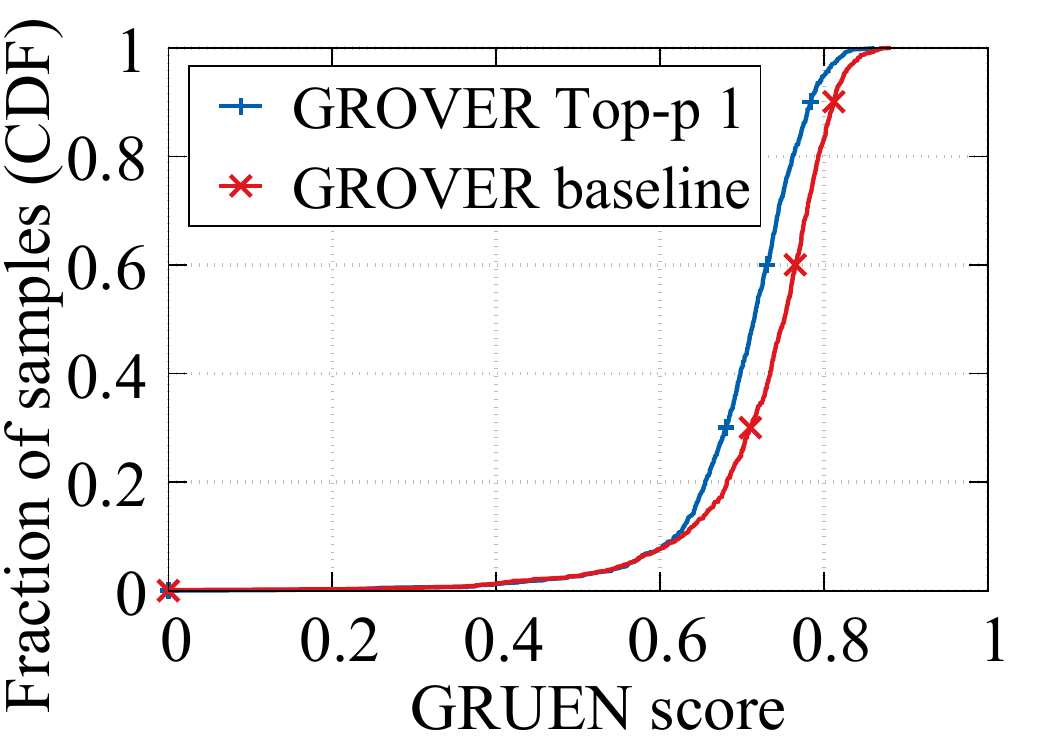}
\caption{The CDF of GRUEN score on \grover's baseline test set (Top-p 0.94 used in Table~\ref{tab:defenses-setup}) and a dataset generated from \grover{} with a decoding strategy of Top-p 1.}
\label{fig:CDF_grover_p1_vs_baseline}
\end{figure}

\para{\textit{Finding 4:} } \textit{Changing the decoding strategy is a simple and effective way to break many defenses.} All defenses, except \fast{}, show significant degradation in $\Delta$R under at least one of the attack strategies, ranging from 13.3\% to 97.6\% degradation. \fast{} does exhibit degradation, but to a lesser extent, compared to the other defenses, \ie degradation in $\Delta$R ranging from 2.9\% to 9.7\%. This suggests that defenses like \fast{} are able to learn more robust features from the text. For \grover{}, \fast{} and \rbert{}, the most effective attack (based on $\Delta$R) happens at a Top-p value of 1.0, which is basically sampling from an untruncated distribution. However, note that the degradation in average GRUEN score is small (<5\%) in these cases. Figure~\ref{fig:CDF_grover_p1_vs_baseline} shows the CDF of the GRUEN scores for text applied to \grover{} at its baseline setting (Top-p 0.94), and when Top-p is 1.0. The two distributions are adjacent,
indicating that using an untruncated distribution is not significantly degrading linguistic quality, thus providing more room for the attacker to fool defenses.

\para{\textit{Finding 5:} }\textit{Classifiers that rely solely on detecting differences in token likelihoods provided by LMs can be easily fooled by changing the decoding strategy.} We observe that \gltrbert{} and \gltrgpt{} break down under text generated from different decoding strategies/parameters. When they encounter text generated using nucleus sampling with a Top-p value of 0.98, the recall of \gltrbert{} and \gltrgpt{} degrades by 90.0\% and 97.6\%, respectively as shown in Table~\ref{tab:decoding_eval_short}. This shows that token likelihood features are highly vulnerable to attacks that change the decoding strategy. With further improvements in decoding strategies that allow more diverse text to be sampled from LMs, such defenses will only be more susceptible to these attacks.

\begin{table}
    \centering
    \setlength{\tabcolsep}{2pt}
    \setlength\extrarowheight{3pt}
    \begin{tabular}{c|c|c|c|c}
    \hline
         \bf \multirow{2}{*}{\bf Defenses}& \bf \makecell{Baseline} & \multicolumn{3}{c}{\bf{Attack: Priming the LM}}  \\
         \cline{3-5}
         &\bf{\#Tokens} &\bf{\#Tokens} & \bf $\Delta$R  & \bf $\Delta$\gruen{} \\
         \cline{3-5}
         \hline
         \gbert{} &1 &0&-47.7& +6.7  \\
         \gltrgpt{} &0&4 &-5.5&-4.9\\
         \gltrbert{} &0&4 & -14.8&-5.0\\
         \grover{} &0& 8 &-1.6&+0.3\\
         \fast{} &0& 12 &-1.8&+0.1\\
         \rbert{} &0& 12 &-1.4&+0.3\\
         \hline
    \end{tabular}
    \caption{Performance of attacks which prime the LM with a different number of priming tokens. Evaluation metrics include Recall (R) of the synthetic class and the average GRUEN (GRN) of synthetic data. We show the percentage change of each evaluation metric ($\Delta$R, $\Delta$GRN) from the baseline performance of each defense on the most effective length of priming tokens. `+' means performance improvement and  `-' means performance degradation.
    }
    \label{tab:priming_eval_short}
\end{table}

\para{Evasion by varying the number of priming tokens}
We test the defenses on text generated using a varying number of priming tokens. 
In Table~\ref{tab:priming_eval_short}, for each defense, we present results for the most effective attack configuration. Similar to the previous adaptive attack (changing decoding strategy), we use $\Delta$R, $\Delta$GRN to measure attack success. Our findings are as follows:

\para{\textit{Finding 6:} }\textit{Defenses trained on conditionally generated text are not able to detect unconditionally generated text.} 
\gbert{} which is originally trained on conditionally generated text (with a single priming token), shows over 47.7\% degradation in $\Delta$R when tested on unconditionally generated text (\ie with 0 priming tokens). 
All the other defenses are trained on unconditionally generated text, and show significantly less degradation, compared to \gbert{}. 
We believe that this is because defenses trained on conditionally generated text learn a narrow distribution of synthetic text. 
We can think of LMs as being in a particular state space at each time-step. The priming tokens might lead the model into a particular state space which the model might not have reached on its own due to those `priming tokens' being low probability tokens, and thus less likely to be sampled in an unconditional setting. Therefore, conditionally trained models might learn a different and narrower distribution of synthetic text and not generalize well to unconditionally generated text. 

\begin{table*}
    \centering
    \setlength{\tabcolsep}{3.5pt}
    \setlength\extrarowheight{2pt}
    \begin{tabular}{c|c|c|c|c|c|c|c|c|c}
    \hline
         \bf \multirow{3}{*}{\bf Defenses} &
         \bf \multirow{3}{*}{\bf \gruen{}-Before} &
         \multicolumn{4}{c|}{\bf \ourattack{}} &
         \multicolumn{2}{c|}{\multirow{2}{*}{\bf Random Perturbations}} & \multicolumn{2}{c}{\multirow{2}{*}{\bf TextFooler}} \\
         \cline{3-6}
         &\multicolumn{1}{c|}{} &  \multicolumn{2}{c|}{\bf \bert{} Backend}& \multicolumn{2}{c|}{\bf GPT2-XL Backend} &\multicolumn{2}{c|}{} &\multicolumn{2}{c}{}\\
         \cline{3-10}
          & &\bf  ER &\bf  \gruen{}-After  &\bf  ER &\bf  \gruen{}-After &\bf  ER &\bf \gruen{}-After &\bf  ER &\bf \gruen{}-After \\
         \cline{2-9}
         \hline
         \gbert{} & 0.652 &1.0 & 0.568&1.0 &0.590&0.7&0.605 &50.4 &  0.480   \\
         \gltrgpt{} & 0.686&91.3 & 0.594 & 74.2&0.633& 61.6&0.642 &99.7& 0.670\\
         \gltrbert{} &0.756 & 44.7&0.654  &45.9 &0.689&32.9 & 0.703&99.8&0.719\\
         \grover{} & 0.734 &59.1&0.647& 43.8&0.678& 30.3&0.691&99.7& 0.720\\
         \fast{} &0.732&24.9&0.646&23.2 &0.676&14.9&0.686 &99.0& 0.689\\
         \rbert{} &0.732&51.4&0.643&44.0 &0.674&26.5&0.687 &99.2 & 0.711 \\
         \hline
    \end{tabular}
    \caption{Attack performance of the three adversarial perturbation methods (\ie \ourattack{}, random perturbations, TextFooler) against the defenses, based on 10 word perturbations. GRN-Before: the average GRUEN of original datasets; GRN-After: the average GRUEN of datasets produced by adversarial perturbation attacks; ER: Evasion Rate (\%).
    }
    \label{tab:adversarial_perturbs_results}
\end{table*}

\para{Evasion by adversarial perturbations.} 
To test \ourattack{}, we use a random sample of 1000 synthetic articles from the original test set of each defense, that were correctly classified. Attack success is measured using the Evasion Rate or ER metric (Section~\ref{adaptive_attack_eval_metric}). Higher ER indicates higher attack success. In addition, our attack requires preservation of semantic content. By design, our perturbation scheme achieves a USE~\cite{cer2018universal} semantic similarity score $\geq$ 0.7, similar to TextFooler. We present our results using a small number of perturbations, \ie $N=10$. The classifier's context window size is 512 tokens, so perturbing 10 words (or less) is a small amount of perturbation. Table~\ref{tab:adversarial_perturbs_results} presents the results for \ourattack{}, and the baseline schemes (TextFooler and random perturbations). Our findings are as follows:




\para{\textit{Finding 7:}} \textit{\ourattack{} can successfully generate adversarial samples without requiring any information about the defense.} From Table~\ref{tab:adversarial_perturbs_results}, we see that \ourattack{} achieves significant evasion rates ranging from 23.2\% to 91.3\% for all the defenses, except \gbert{}. \ourattack{} with BERT and GPT2-XL as the backend also outperforms the random perturbation attack setting for all defenses. While TextFooler outperforms \ourattack{} for all defenses, it is important to note that TextFooler makes a large number of queries 
to the defense model to craft more effective samples whereas \ourattack{} does not require queries to the model.
Also, the average GRUEN scores of samples with random perturbations are comparable to the GRUEN scores of \ourattack{} with GPT2-XL backend---across all defenses, the average absolute difference between GRUEN scores is only 0.014. 

\begin{figure}[H]
\centering
\includegraphics[width=0.9\columnwidth]{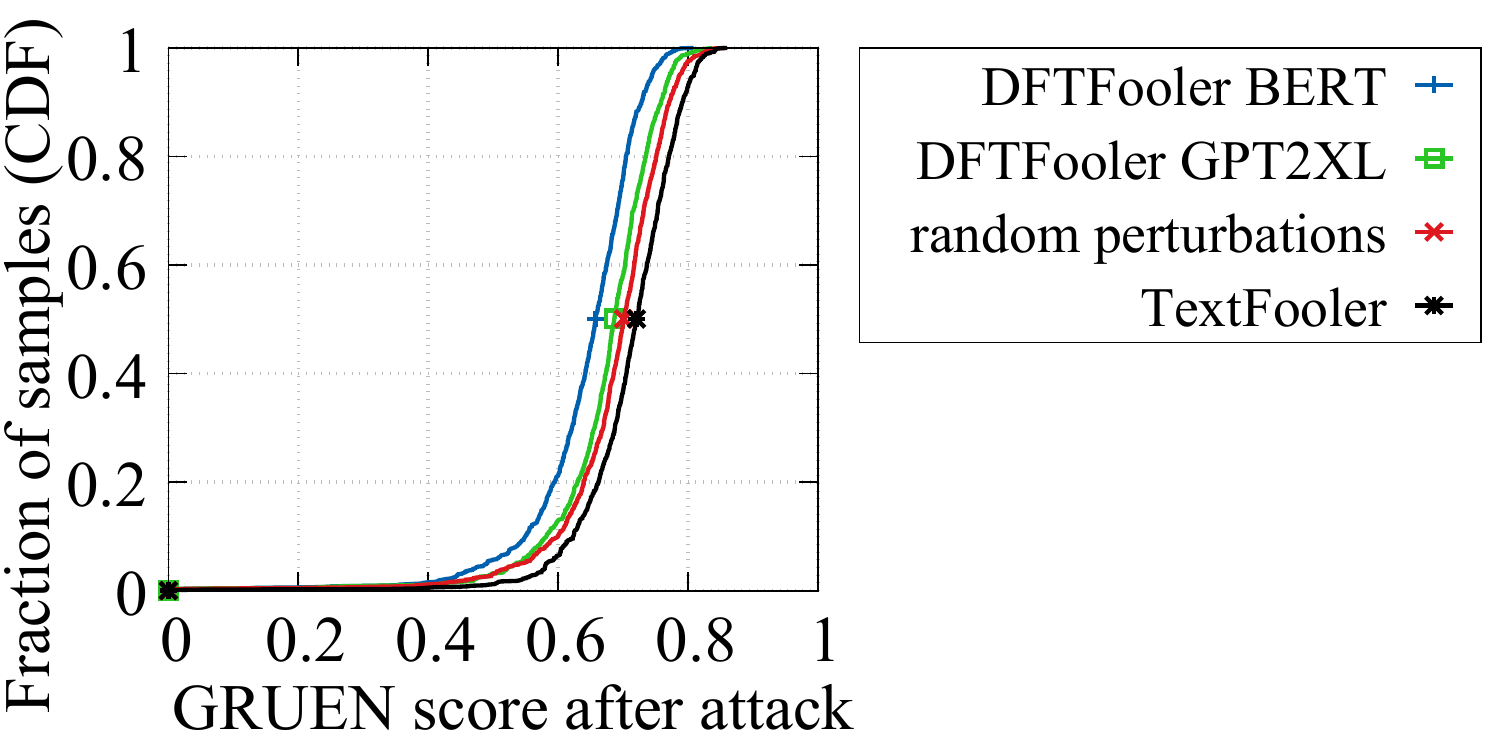}
\caption{The CDF of GRUEN score on perturbed text produced by different adversarial perturbation methods, 
\ie \ourattack{}, random perturbations, and TextFooler, based on attacking \rbert{} (in Table~\ref{tab:adversarial_perturbs_results}).}
\label{fig:GRUEN_CDF_5curves}
\end{figure}

\para{\textit{Finding 8:}} \textit{\ourattack{} using a bidirectional LM as backend, provides more effective adversarial samples.}
In Table~\ref{tab:adversarial_perturbs_results}, \ourattack{} has higher ER when it uses \bert{} as the backend model compared to GPT2-XL, against 4 out of 6 defenses. 
More specifically, \ourattack{} with BERT backend shows percentage increase in ER ranging from 35.8\% to 95.0\% compared to random perturbations. 
In other words, using bidirectional context to compute token probabilities, provides better estimation of important words to be replaced.
That said, we observe a slightly higher hit on GRUEN score for \ourattack{} with the BERT backend compared to when GPT2-XL is used as the backend. We suspect that this is because the GRUEN score is computed, in part, by a pre-trained BERT model~\cite{zhu2020gruen}.
There is likely an overlap between the words \ourattack{} chooses to perturb and the words the GRUEN metric assigns more importance to for computing sub-scores that utilize a pre-trained BERT. Figure~\ref{fig:GRUEN_CDF_5curves} (Appendix) shows the GRUEN score distribution (CDF) for the different attacks.

 

\para{\textit{Finding 9:}} \textit{Increasing the number of word perturbations will improve the Evasion Rate but degrade text quality.}  
The attack performance of \ourattack{} and random perturbations in Table~\ref{tab:adversarial_perturbs_results} are obtained with 10 word perturbations. To understand the impact of the number of perturbations, we experiment with a different number of word perturbations for \gbert{} and \fast{} and present the results in Figure~\ref{fig:word_perturb_plot} in Appendix. We observe a clear trend that for both \ourattack{} and random perturbations, increasing the number of perturbations will result in a higher evasion rate, but at the cost of increased degradation in the GRUEN score. 

    \subsection{Towards Adversarial Robustness} 
\label{sec:towards_adv_robustness}
\para{Understanding robustness when using semantic features.}\label{DistilFAST}
Among all the defenses, \fast{} has held up more consistently against the different attacks, and also generalizes well to content in the wild. But it is still unclear what aspect of FAST contributes to its robustness. Our hypothesis is that \fast{}'s performance can be attributed to the use of semantic features based on entities mentioned in the article. \fast{} models the factual structure of the article by tracking the consistency of named-entities mentioned in it. To validate this hypothesis, we analyze \fast{} in more detail.


\begin{figure}[!ht]
\centering
\includegraphics[width=1\columnwidth]{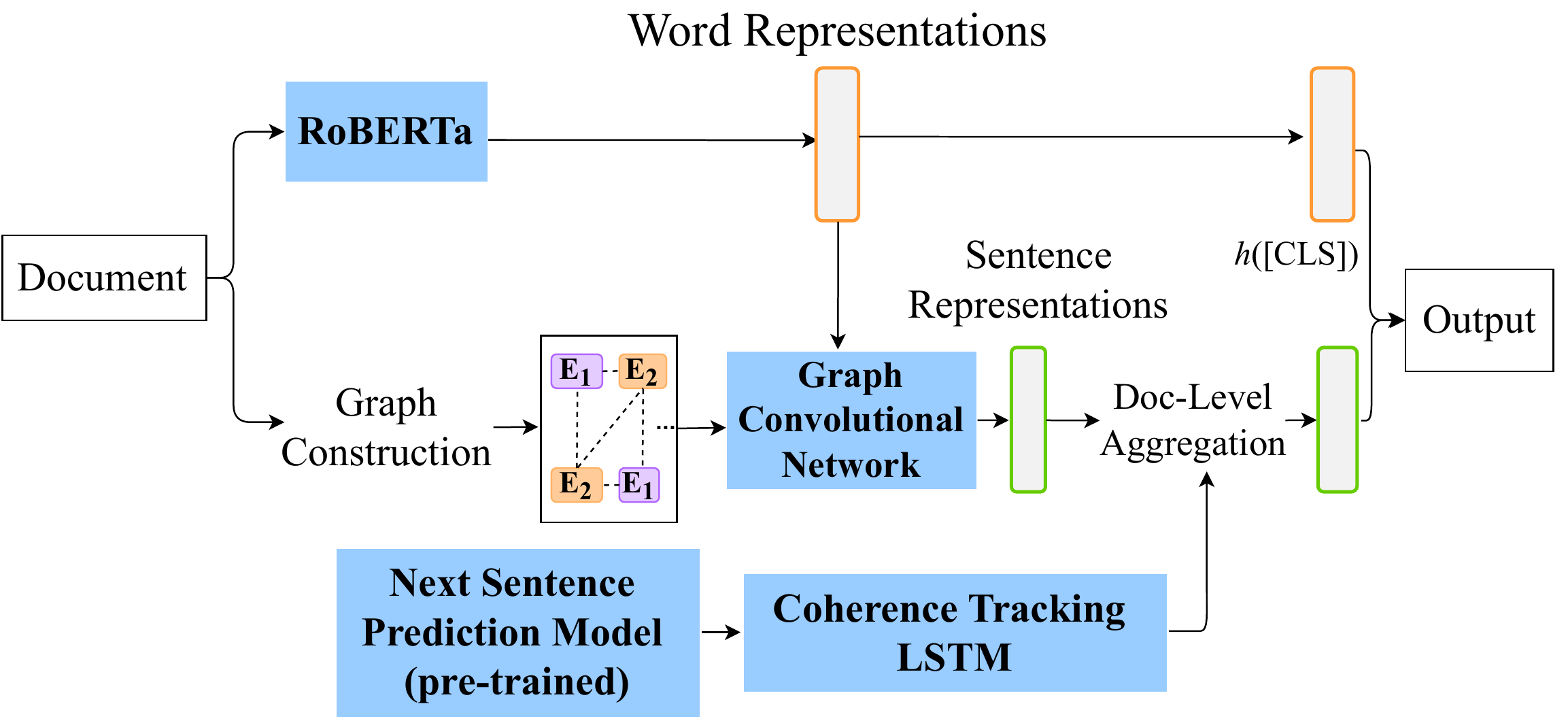}
\caption{An overview of the FAST pipeline.}
\label{fig:fast_method_illustration}
\end{figure}

FAST has complex internals. As shown in Figure~\ref{fig:fast_method_illustration}, FAST comprises 4 main components: A RoBERTa-based feature extractor, a Multi-layer GCN, a Next Sentence Prediction (NSP) model and a coherence tracking LSTM. Taking a document as the input, \fast{} first learns contextual semantic representations for words via the RoBERTa language model. Next, a graph containing nodes representing entities in the text is created. The contextual word embeddings from RoBERTa are concatenated with Wikipedia2vec~\cite{yamada2020wikipedia2vec}
entity representations to form the embedding for the entity graph. This graph embedding is then fed to a multi-layer GCN to obtain graph-enhanced sentence embeddings, which are then fed to an LSTM for coherence tracking. A Next Sentence Prediction (NSP) model is used to calculate the contextual coherence score for each neighbouring sentence pair. The NSP scores are then used to compute a document-level representation from the LSTM outputs. Finally, the RoBERTa embeddings are concatenated with the document-level representation and fed to a classification layer.

To better understand \fast{}'s superior performance, it is important to break down its complexity. We do so by running multiple ablation experiments. Models for the ablation studies are trained using the same data used to train \fast{}.


\textit{Ablation experiment \#1: \rbert{}.} We begin by considering a defense that only uses the RoBERTa language model. This is the same \rbert{} that has been evaluated in the previous sections of the paper.
\rbert{} remains robust in several attacks, compared to the other defenses, but performs worse than \fast{} in two settings: (1) when under attack by \ourattack{} (Table~\ref{tab:adversarial_perturbs_results}), and (2) when under attack by varying the decoding strategy (Table~\ref{tab:decoding_eval_short}). For example, \rbert{} suffers a degradation in Recall of 22.0\% compared to \fast{} which deteriorates only by 9.7\%, when it encounters text generated using Top-p 1.0 decoding. This indicates that RoBERTa is not the main source for the robustness of \fast{}.



\textit{Ablation experiment \#2: DistilFAST.} Next, we test whether the semantic features, \ie features from the entity network extracted by the GCN are the source of \fast's robustness. To do so, we create a ``distilled'' version of \fast{}, called DistilFAST, by removing the NSP task, the LSTM coherence tracker, and the Wikipedia embeddings for the GCN from \fast's pipeline. As a result, we are left with the RoBERTa model and the GCN. To create the document-level representation, we compute the element-wise sum of the sentence-level representations obtained from the GCN.
To test robustness of DistilFAST, we evaluate it against : \textit{(1)} adaptive attacks changing the generation process, \textit{(2)} adversarial inputs based on \ourattack{} and random perturbations, and (3) \textit{In-the-wild} datasets. \textit{If DistilFAST performs similar or better than \fast{}, it would suggest that use of entity-based semantic features is the key enabler for \fast{}'s better generalization and robustness.}

\begin{table}[t]
    \centering
    \setlength\extrarowheight{2.5pt}
    \begin{tabular}{P{2.8cm}|P{1.1cm}|P{1.1cm}|P{1.1cm}}
    \hline
    \bf \multirow{3}{*}{\bf{Attack Strategy}} & \multicolumn{3}{c}{\bf Detection Performance (Recall)}\\
    \cline{2-4} &\multirow{2}{*}{\bf{\makecell{Distil\\FAST}}} & \multirow{2}{*}{\bf \fast{}} & \multirow{2}{*}{\bf \makecell{\makecell{RoBERTa-\\Defense}}} \\
    &&&\\
    \cline{2-4}
    \hline
    Top-p 1.0 &80.1 &81.6&71.5\\
    Top-k 80 &88.7 &87.8&90.9\\
    Priming Tokens 12 &89.2&88.8&90.4\\
    \hline
    \end{tabular}
    \caption{Detection performance (Recall) of DistilFAST, \fast{} and \rbert{} on attacks which change the decoding strategy used by the LM or prime it with varying numbers of priming tokens. 
    }
    \label{distilfast_decoding_priming_recall}
\end{table}

\textit{DistilFAST against attacks changing the generation process.} Table~\ref{distilfast_decoding_priming_recall} shows the results. We consider attacks that change the text decoding strategy and the number of priming tokens.
We test DistilFAST on the most effective attack configurations against \fast{} (from Tables~\ref{tab:decoding_eval_short} and~\ref{tab:priming_eval_short}). DistilFAST achieves a similar Recall as \fast{} when changing the decoding strategy to use Top-p 1.0 setting. In the other two strategies (using Top-k 80 and changing the number of priming tokens), DistilFAST even slightly outperforms \fast{}. We also show results for \rbert{}, which does not exhibit similar performance as \fast{} in one of the settings. 
\textit{DistilFAST against \ourattack{} and random perturbations.} When the \ourattack{} attack is applied to DistilFAST, we observe a 10.4\% and 9.4\% reduction in ER, compared to \fast{}, when using the BERT and GPT2 backend for \ourattack{}, respectively. Similarly, against random perturbations, we observe a 13.4\% reduction in ER, compared to \fast{}. DistilFAST is able to achieve better adversarial robustness than \fast{} against our adversarial perturbations.



\begin{table}
    \centering
    \setlength\extrarowheight{3pt}
    \begin{tabular}{P{2.5cm}|P{1.2cm}|P{1.2cm}|P{1.5cm}} 
     \hline
    \bf \multirow{3}{*}{\bf{Datasets}} & \multicolumn{3}{c}{\bf Detection Performance (F1)}\\
    \cline{2-4} &\multirow{2}{*}{\bf{\makecell{Distil\\FAST}}} & \multirow{2}{*}{\bf \fast{}} & \multirow{2}{*}{\bf \makecell{\makecell{RoBERTa-\\Defense}}} \\
    &&&\\
    \hline
    AI-Writer &94.1&94.9&92.1\\  
    ArticleForge &88.2&88.5&87.4\\
    \hline
    \end{tabular}
    \caption{Detection performance (F1) of DistilFAST, \fast{} and \rbert{} on AI-Writer and ArticleForge. 
    }
    \label{distilfast_wild_recall}
\end{table}

\textit{DistilFAST against In-the-wild datasets.}
We evaluate DistilFAST against the \textit{In-the-wild} datasets from the news domain, \ie ArticleForge and AI-Writer). Detection performance results are presented in Table~\ref{distilfast_wild_recall}. DistilFAST performs similar to \fast{}, suggesting that entity-based semantic features can improve generalization performance.


Our analysis leads to the following key finding:

\para{\textbf{Finding 10}: Semantic features that capture the factual structure of the text, \ie entity-level features, provides robustness against adaptive attacks and better generalization performance.}


\begin{table}
    \centering
    \setlength{\tabcolsep}{3pt}
    \setlength\extrarowheight{3pt}
    \begin{tabular}{c|c|c|c|c}
    \hline
    \bf \multirow{4}{*}{\bf{Defenses}} & \multicolumn{4}{c}{\multirow{2}{*}{\bf{\makecell{Detection Performance(Recall) \\ with Adversarial Training}}}}\\ \\
    \cline{2-5} &\multirow{2}{*}{\bf{\makecell{  Adaptive \\Attack}}} & \multirow{2}{*}{\bf\makecell{Recall \\ Before}} & \multirow{2}{*}{\bf \makecell{Recall \\After}} &\multirow{2}{*}{\bf $\Delta$R}\\
    &&&&\\
    \hline
        \gbert{}&Priming Token 0&44.9&85.8&+91.1\\
        \gltrgpt{} &Top-k 160&42.8& 99.2&+131.8 \\
        \grover{}&Top-p 1.0&58.7&77.5&+32.0\\
         \hline
    \end{tabular}
    \caption{ Detection performance (Recall) of \gbert{}, \gltrgpt{}, and \grover{} when fine-tuned to their most effective adaptive attack setting as indicated. $\Delta$R is percentage change of Recall from Recall before training to Recall after training.
    }
    \label{tab:fine_tune_decoding}
\end{table}


\para{Adversarial training to enable robustness against adaptive attacks.}
We investigate whether a defender can recover from an adaptive attack via \textit{adversarial training}, \ie by training a defense on a set of known adversarial samples to build resilience against similar adversarial samples. We start by studying recovery from adaptive attacks that change the text generation process. We fine-tune \gbert{}, \grover{} and \gltrgpt{}~\footnote{Since \gltrgpt{} is not DNN-based, we instead train a new \gltrgpt{} model from scratch with samples generated under the attack setting.} on new samples generated from their most effective adaptive attack setting (Tables~\ref{tab:decoding_eval_short} and~\ref{tab:priming_eval_short}). We use 1,000 new articles each for both the synthetic and real class for adversarial training. We then evaluate the adversarially trained models against their original adaptive attack dataset. As shown in Table~\ref{tab:fine_tune_decoding}, we observe that the fine-tuned \gbert{}, \gltrgpt{} and \grover{} achieve 91.1\%, 131.8\% and 32.0\% increase in $\Delta$R, respectively. Therefore, the fine-tuned models are able to recover from the attack.

Next, we explore adversarial training to recover from our \ourattack{} attack. We use \rbert{} for this experiment. From our dataset of 1,000 adversarial samples from \ourattack{}, we use a random set of 500 samples for adversarial fine-tuning, and test the adversarially trained \rbert{} on the remaining 500 adversarial samples. \ourattack{} becomes ineffective as its evasion rate drops from 51.4\% on \rbert{} (in Table~\ref{tab:adversarial_perturbs_results}) to 1.6\% on the adversarially fine-tuned \rbert{}.

Fine-tuning a defense towards a specific attack is effective, but the defender may have to frequently adapt their defenses against newer adaptive strategies or their variants. On the other hand, by using robust semantic features, one can potentially build in resilience in an attack-agnostic way.

\section{Discussion}
\label{sec:discussion}
\para{Deepfake text detection vs. deepfake image detection.} 
Deepfake image detection and deepfake text detection both aim to detect synthetic content generated by deep generative models. That said, these two fields hardly share the same set of technical methodologies in generating and detecting synthetic content due to the discrete nature of text. Similar to the text domain, researchers have also demonstrated impressive performance in detecting deepfake images~\cite{frank2020leveraging, wang2020cnn}. This has prompted work investigating the adversarial robustness of deepfake image detectors~\cite{carlini2020evading, gandhi2020adversarial, cao2021understanding}. In a similar way, our work is the first to systematically explore real-world performance and adversarial robustness of deepfake text detectors. While the image domain has primarily focused on adversarial perturbations, our work also explores low-cost evasion strategies that change the content generation process. Note that the adversarial attacks studied for the image domain are not directly applicable to the text domain. If we compare detection schemes in the two domains, more progress has been made on understanding the use of semantic features for deepfake image detection~\cite{li2018ictu, li2018exposing, ciftci2020fakecatcher, matern2019exploiting}. An example is a method that looks for eye blinking artifacts~\cite{jung2020deepvision} to detect deepfake images. However, in the text domain, \fast{} is the only approach that leverages semantic features (entity-based), and there is scope for more work to be done in this direction.

\para{Further exploring semantic layer methods for synthetic text detection.}
Producing semantically consistent text is still a challenging task for language models~\cite{dou2021scarecrow}. 
Therefore, we expect to find differences in the semantic information embedded in synthetic text when compared to real text. In this study, we show that \fast{} leverages such imperfections in the semantic structure of the text to differentiate between synthetic and real text. Using semantic information will also raise the cost of producing synthetic text that can bypass such defenses. In practice, the attacker may want to preserve the semantic content, \eg spreading vaccine disinformation, while creating evasive samples, thus making it harder to evade detection.
One direction for future work is to leverage \textit{knowledge graphs} to extract richer semantic features. 

\para{Ethics.}
\noindent Our work involved collecting synthetic text data from \textit{text-generation-as-a-service} platforms and from Internet forums. We spent \$586 to collect the articles from the services. All the services we study claim that the synthetic articles can be used for white hat SEO. Regardless of the legal status of these services, the benefits gained from understanding how well the state-of-the-art defenses perform on synthetic text generated from them outweighs the potential harms arising from injecting money into these services.
Our work proposes as well as evaluates existing defenses against adversarial inputs. This was done in a controlled lab setting, and no deployed models were attacked in this process.


\section{Conclusion}
\noindent To the best of our knowledge, this work presents the first systematic evaluation of deepfake text defenses to assess their real-world applicability. We evaluated state-of-the-art synthetic text defenses on real-world datasets and against adaptive attackers. We find that open-domain detection schemes fail to generalize to 
\textit{In-the-wild} synthetic text, and that most defenses are not robust under adversarial settings. We also presented \ourattack{}, a novel adversarial sample crafting scheme, that can degrade the performance of existing defenses without requiring any queries to the victim model nor a surrogate classifier. Our detailed analysis of the most robust defense (FAST) indicates that utilizing semantic information in the text samples can lead to better robustness and generalization performance in the wild.

\section*{Acknowledgment}

\noindent This research was supported by startup funds from Virginia Tech. Zain Sarwar was funded by the Higher Education Commission (HEC) Pakistan National Center for Cyber Security (NCCS) grant at LUMS. We thank Neal Mangaokar for discussions of text decoding strategies and their impact on text generation.

\bibliographystyle{IEEEtran}
\bibliography{main}
\section*{Appendix}

\subsection{Metrics for Evaluating Linguistic Quality}\label{other_text_quality_metrics}
\noindent Besides GRUEN, we surveyed other text quality metrics and categorized them into the following groups: word-based metrics, embedding-based metrics, training-based metrics, and dialog-based metrics. Here we explain the main features of each metric category and their limitations given our usage scenario. Both word-based and embedding-based metrics require human reference text for evaluating text quality. Word-based metrics compute text quality based on the word or n-gram overlap between the evaluated text and the reference text, e.g., BLEU~\cite{papineni2002bleu}, METEOR~\cite{banerjee2005meteor}, NIST~\cite{doddington2002automatic}, and ROUGE~\cite{lin2004rouge}. Word-based metrics rely largely on word-level matches. Such similarities can be better captured by word embeddings such as Word2Vec~\cite{mikolov2013efficient} and GloVe~\cite{pennington2014glove}. Thus, an alternative to matching words is to compare the similarity between the embeddings of words in the evaluated text and the reference, e.g., Greedy Matching~\cite{rus2012optimal} and Embedding Average metric~\cite{Landauer1997AST}. 
The main limitation of these metrics is that all of them require references from the real class, which are not available in our case. Instead of comparing with human generated gold standard references, training-based metrics contain learnable components that are trained specifically for the task of automatic evaluation, e.g., ADEM~\cite{lowe2017towards}. However, training this model requires human annotations which were not available to us. Other text quality metrics such as GRADE~\cite{huang2020grade} and USR~\cite{mehri2020usr} were designed for evaluating the quality of synthetic dialogs, and are difficult to transfer across usage domains. Besides automatic evaluation metrics, another way to evaluate text quality is to conduct a human study to annotate the quality of documents. However, given the large number of datasets we evaluated, this was not realistic for us to do.

\subsection{A Detailed Description of GRUEN Score}\label{gruen_subscores}

\noindent GRUEN, proposed by Zhu \etal{}~\cite{zhu2020gruen}, is an unsupervised and reference-less text quality metric. Zhu \etal{} show that GRUEN is more correlated with human judgement of text quality than any other existing metric. The GRUEN score of an article is computed by aggregating the following sub-scores: 

\para{Grammaticality.} This is computed by combining two sub-scores: Perplexity and grammar acceptance. Perplexity is computed using a \bert{} model whereas the grammar acceptance score is computed by fine-tuning a \bert{} on the CoLA dataset~\cite{warstadt2019neural} which contains labelled examples of grammatically correct and incorrect sentences.

\para{Non-redundancy.} This metric computes whether a document contains excessively repeated sentences, phrases and instances where proper nouns were used instead of pronouns. This is done by computing four inter-sentence syntactic features: length of the longest common substring, count of the longest common words, edit distance and the number of common words in a document. 

\para{Focus.} This score looks at the semantic similarity between adjacent sentences as a measure of discourse focus. It is computed via the Word Mover Similarity~\cite{pmlr-v37-kusnerb15} for adjacent sentences.  

\para{Structure and coherence.} This is calculated by computing the loss on the Sentence-Order-Prediction~\cite{lan2020albert} task as it models the inter-sentence coherence in a document. In the code provided by the authors of the GRUEN metric, this component was not included.

\subsection{Linguistic Quality of Synthetic Text in the Wild}\label{sec:wild_text_quality}
\begin{figure}[H]
\centering
\includegraphics[width=0.9\columnwidth]{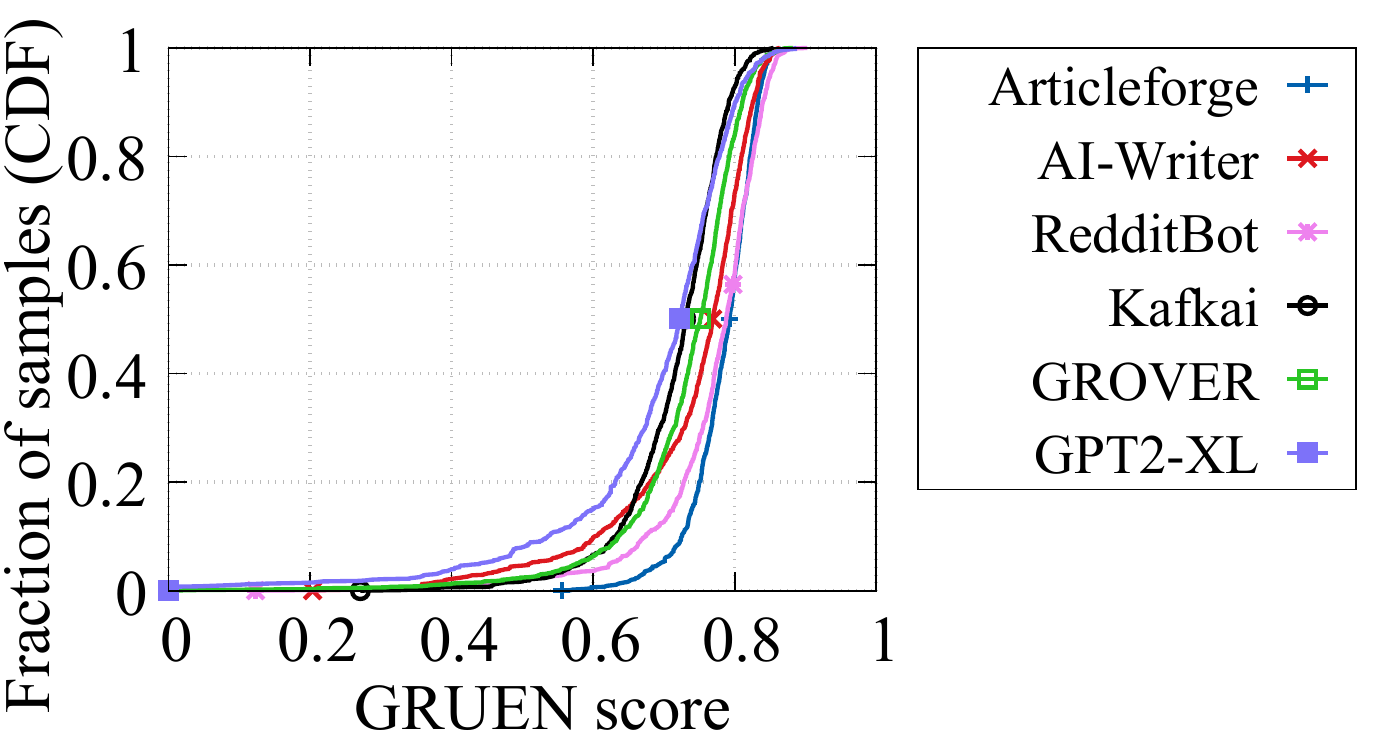}
\caption{The CDF of GRUEN on the 4 \textit{In-the-wild} datasets and 2 datasets generated from \grover{} and GPT2-XL.
}
\label{fig:grune_score_6data}
\end{figure}

\noindent We use the GRUEN metric to evaluate linguistic quality. Figure~\ref{fig:grune_score_6data} shows the CDF of GRUEN scores for \textit{In-the-wild} synthetic samples, compared with synthetic text produced by the research community, which includes GPT2-XL and \grover{}. We can see that data in the wild is comparable or better than synthetic text produced by the research community. 

\subsection{Other Defenses}\label{sec:other_defenses}
\noindent There are a few other defenses that are not considered in our study.
Yao \etal{}~\cite{yao-2017-dnnreviews} in 2017, proposed a method to detect LSTM-generated synthetic reviews (restaurant reviews targeting Yelp). While the proposed method works well against LSTM-generated text, synthetic text generation has advanced significantly since Transformers were introduced. Yao's method uses an LSTM-based supervised approach to detect synthetic text. We omit this defense because our preliminary evaluation of this approach on synthetic text produced by Transformers yielded unsatisfactory results. We upgraded Yao's approach to use a Transformer-based classifier (instead of an LSTM model), and trained the model on 5000 real articles from the RealNews dataset, and 5000 articles produced by \grover{}. Unfortunately, Yao's approach only achieves an F1-score of 68\% in detecting \grover{} generated text.

Adelani \etal{}~\cite{adelani2019generating} present several classifiers, \eg \gltr{}, \grover{} and an OpenAI GPT-2 based Detector~\cite{solaiman2019release} to detect synthetic reviews generated using GPT2-Small. Fagni \etal{}~\cite{fagni2021tweepfake} develop several classifiers based on Markov Chains, RNNs, LSTMs, and GPT-2 to detect synthetic tweets on Twitter. We do not study both approaches because many of the state-of-the-art methods such as \grover{}, \gltr{} and other Transformer-based models are already considered in our work. Moreover, based on our preliminary investigation of Yao \etal{}'s work, we found that RNN/LSTM-based models are not promising approaches to detect synthetic text produced by advanced models like Transformers. 

\subsection{Applying Grid Search for Training GLTR}\label{sec:gltr_grid_search}
\noindent 
To tune the hyperparameters of GLTR defenses (based on logistic regression classifier), we apply grid search to the training process of the GLTR defenses, which includes \gltrgpt{} and \gltrbert{}. We use \textit{GridSearchCV}~\cite{gridsearch_cv} which is built into scikit-learn to select the model.
Given a set of parameter values, GridSearchCV can exhaustively consider all parameter combinations, fit the model on the training set, and select the best model. 
To train \gltr{} defenses, we apply grid search on the following hyperparameters of the logistic regression classifier, \ie ``solver'', ``penalty'' and ``C''. Different choices of `solver'', ``penalty'' and ``C'' can result in differences in model performance~\cite{logistic_regression_sk}.
We use GridSearchCV to loop through the following parameter values---[`newton-cg', `lbfgs', `liblinear'], [`l2'], [100, 10, 1.0, 0.1, 0.01] for ``solver'', ``penalty'' and ``C'', respectively.

\subsection{Details of Fine-tuning Experiments}\label{sec:exp_details}
\para{\gbert{} fine-tuning experiments with In-the-wild samples.}
To improve \gbert{}'s detection performance on \textit{In-the-wild} datasets, we fine-tune \gbert{} with a limited set of \textit{In-the-wild} samples (\ie 10, 50 and 100 samples)  in Section~\ref{sec:improve-in-the-wild}. We follow the general guidelines of transfer learning to fine-tune \gbert{}~\footnote{\url{https://huggingface.co/docs/transformers/training\#finetune-a-pretrained-model}}.
We set batch size as 4, and fine-tune the \gbert{} model for 8 epochs. 
While doing this experiment, 
we encountered a known problem of instability of fine-tuning BERT on small datasets~\cite{zhang2020revisiting}. To overcome this problem, we employ the revitalization strategy proposed by Zhang et al.~\cite{zhang2020revisiting}. We also tune certain training parameters to improve our results. Specifically, we use the ``\textit{adamw\_torch}'' optimizer, and set the ``\textit{weight\_decay}'' and ``\textit{warmup\_ratio}'' to 10e-5 and 0.3, respectively.

\para{\grover{} fine-tuning experiments with In-the-wild samples.}
To improve \grover{}'s detection performance on \textit{In-the-wild} datasets, we fine-tune \grover{} with a limited set of \textit{In-the-wild} samples (10, 50 and 100 samples). We use a batch size of 4, and fine-tune the \grover{} model for 3 epochs. 


\begin{table*}[!htp]
    \centering
    \setlength\extrarowheight{3pt}
    \begin{tabular}{cc}
    \hline
    \textbf{Tools} & \textbf{Websites} \\
    \hline
    ArticleForge & \makecell{politico.com, usatoday.com, deseretnews.com, hollywoodreporter.com, theatlantic.com,\\
    nbcphiladelphia.com, reuters.com, reuters.com, dailymail.co.uk, theguardian.com} \\  
    \hline
    AI-Writer & \makecell{arabnews.com, bbc.com, dailymail.co.uk, dailytimes.com.pk, dawn.com,\\
    deseret.com, esquire.com, gizmodo.com, hollywoodreporter.com, mashable.com,\\
    nbcphiladelphia.com, nj.com, politico.com, reuters.com, theatlantic.com, theglobeandmail.com,\\
    thenorthernecho.co.uk, thisismoney.co.uk, usatoday.com, thedailystar.net} \\
    \hline
    \end{tabular}
    \caption{Websites used to scrape real news articles for ArticleForge and AI-Writer datasets.}
    \label{tab:ai_writer_aforge_websites_appendix}
\end{table*}

\begin{table*}[!htp]
    \centering
    \setlength\extrarowheight{3pt}
    \setlength{\tabcolsep}{1pt}
    \begin{tabular}{P{1.8cm}|P{0.8cm}|P{0.8cm}|P{0.8cm}|P{0.8cm}|P{0.8cm}|P{0.8cm}|P{0.7cm}|P{0.7cm}|P{0.7cm}|P{0.7cm}|P{0.7cm}|P{0.7cm}|P{0.7cm}|P{0.7cm}|P{0.7cm}|P{0.8cm}|P{0.8cm}|P{0.8cm}}
    \hline
         \multirow{2}{*}{\bf Datasets} & \multicolumn{3}{c|}{\bf \gbert{}} & \multicolumn{3}{c|}{\bf \gltrgpt{}} & \multicolumn{3}{c|}{\bf \gltrbert{}} & \multicolumn{3}{c|}{\bf \grover{}} & \multicolumn{3}{c|}{\bf \fast{}}& \multicolumn{3}{c}{\bf \makecell{RoBERTa-Defense}}  \\
         \cline{2-19}
         & \bf F1 &\bf  P &\bf  R & \bf F1 &\bf  P &\bf  R &\bf  F1 &\bf  P &\bf  R &\bf  F1 &\bf  P &\bf  R & \bf F1 &\bf  P &\bf  R & \bf F1 &\bf  P &\bf  R \\
         \cline{2-19}
         \hline
         AI-Writer & 28.8& 73.1 & 17.9& 1.6& 100& 0.8&64.8&77.5&55.7&87.6 &78.4 & 99.3 & 94.9 & 91.0 & 99.2 &92.1&86.9&98.1 \\
         ArticleForge &19.7&59.0&11.8 &44.1 &97.3 & 28.5&85.6&76.3&97.6&76.9&62.6 &99.8 & 88.5 & 84.7 & 92.7 &87.4&80.1&96.3\\
         Kafkai &65.9&90.0 &52.0 &1.0&62.5&0.5&41.4 &57.6 &32.3  &--&--&--&--&--&--&--&--&--\\
         RedditBot &14.1 &25.8 &9.7 &61.5& 99&44.6 &83.4&72.8&97.5  &--&--&--&--&--&--&--&--&--\\
         \hline
    \end{tabular}
    \caption{Performance of the defenses on the \textit{In-the-wild} datasets. We present F1 score (F1), Precision (P), and Recall (R) of the synthetic class in percentages. 
    We did not test \grover{}, \fast{} and \rbert{} on non-news domain datasets, which includes Kafkai and RedditBot. This is because these defenses are only trained for the news domain.
    }
    \label{tab:eval_in_the_wild_long}
\end{table*}

\subsection{More Details of the DFTFooler Pipeline}\label{sec:dft_details}
\para{Identifying a set of important words to be replaced via a LM.} \ourattack{} scans a given article to choose the top $N$ most confidently predicted words according to the chosen LM, for replacement. 
For example, GPT-2 is a standard left-to-right language model, and after tokenizing the articles into a sequence of tokens \{$x_{1}$, $x_{2}$, ..., $x_{i}$\}, GPT-2 can compute the prediction probability of token $x_{i}$, using Equation~\ref{eqn: language_model}, \ie $p(x_i|x_0,x_1,...,x_{i-1})$. For simplicity, assume that each token is a word in the article. At each step in the sequence, a word is assigned a rank among all the words in the vocabulary based on its prediction probability score (higher probability leads to higher rank). In practice, a word can be tokenized into multiple tokens. For words that are tokenized into multiple tokens by the tokenizer, \ourattack{} uses the probability score of the word's first subtoken as the word's probability score. This way, each word in the sequence is assigned a rank based on this probability score. We eventually choose the set of top N most highly ranked words.

\subsection{List of Real News Websites}\label{sec:list_of_news_websites} 
\noindent As explained in Sec ~\ref{sec:in_the_wild_datasets}, our \textit{In-the-wild} dataset contained an equal number of real and fake articles. For generating news articles, we used two text generation services, namely AI-Writer and ArticleForge. 

ArticleForge can generate fake articles with a set of provided keywords. In this case, we collected 1000 real news articles from 10 news websites, and used keywords from them to generate 1000 fake news articles. 

On the other hand, AI-Writer requires a title to generate an article. Similar to the method used for ArticleForge, we scraped 1000 news articles from a list of 20 news websites, and used their titles to generate synthetic articles. In both cases, the list of news websites are listed in Table ~\ref{tab:ai_writer_aforge_websites_appendix}.

\begin{figure}[H]
\centering
\includegraphics[width=0.9\columnwidth]{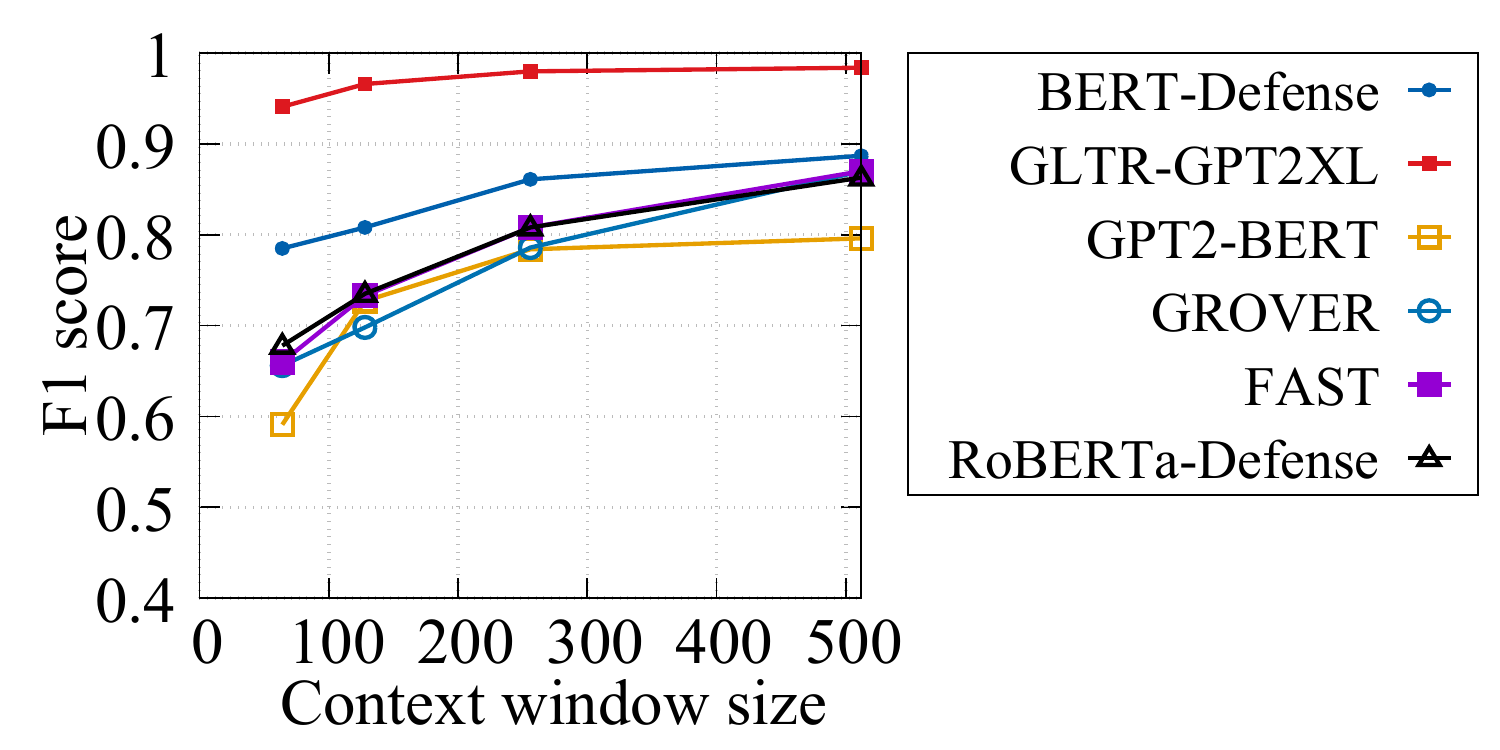}
 \caption{Detection performances of the defenses with different context window sizes, \ie 64, 128, 256, 512 tokens.
}
\label{fig:baseline_token_length}
\end{figure}


\begin{figure}[H]
\centering
\begin{subfigure}[t]{0.48\columnwidth}
  \includegraphics[width=\columnwidth]{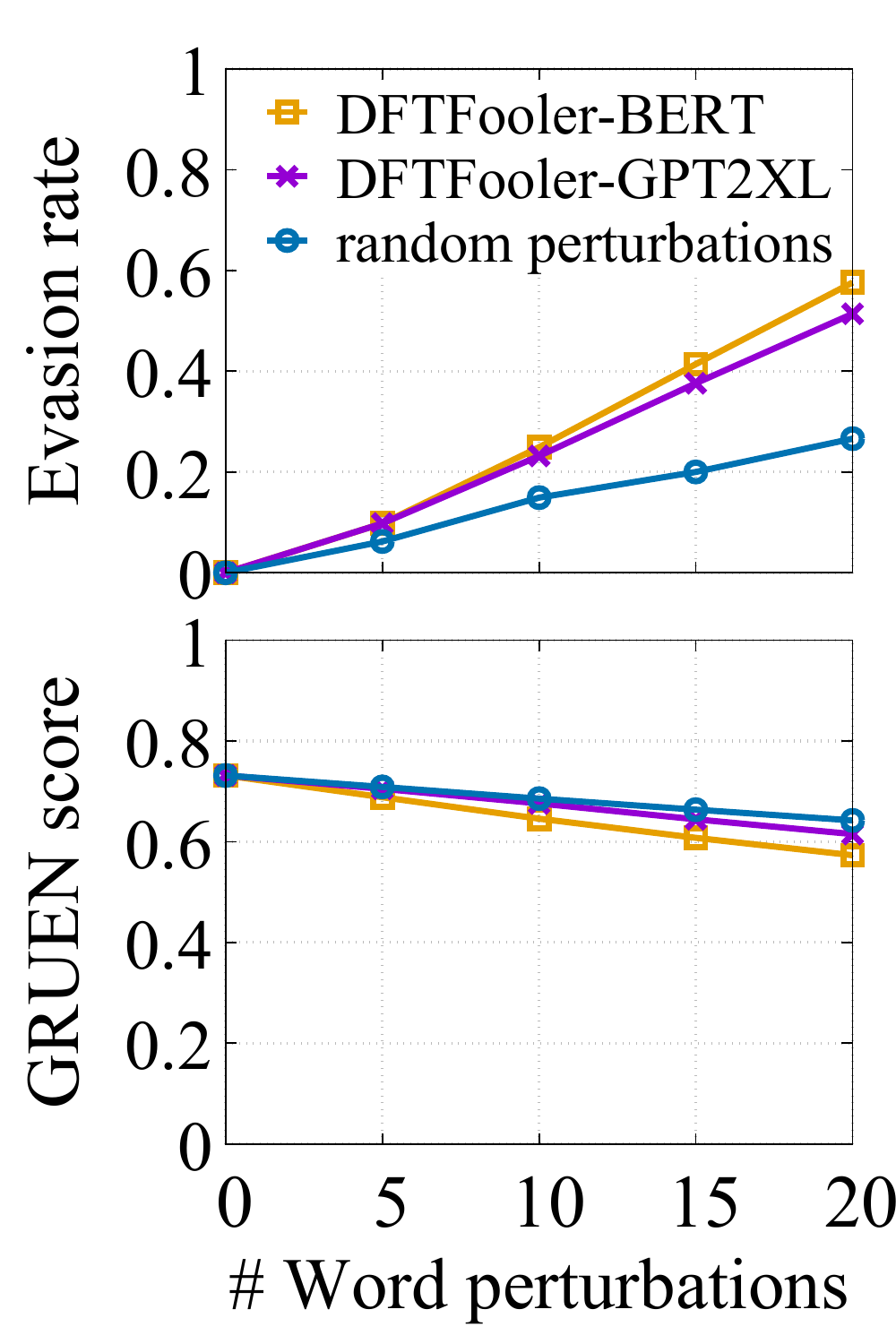}
\caption{Attacking \fast{}.}
\label{fig:fast_perturb}
\end{subfigure}
\hfill
\begin{subfigure}[t]{0.48\columnwidth}
  \includegraphics[width=\columnwidth]{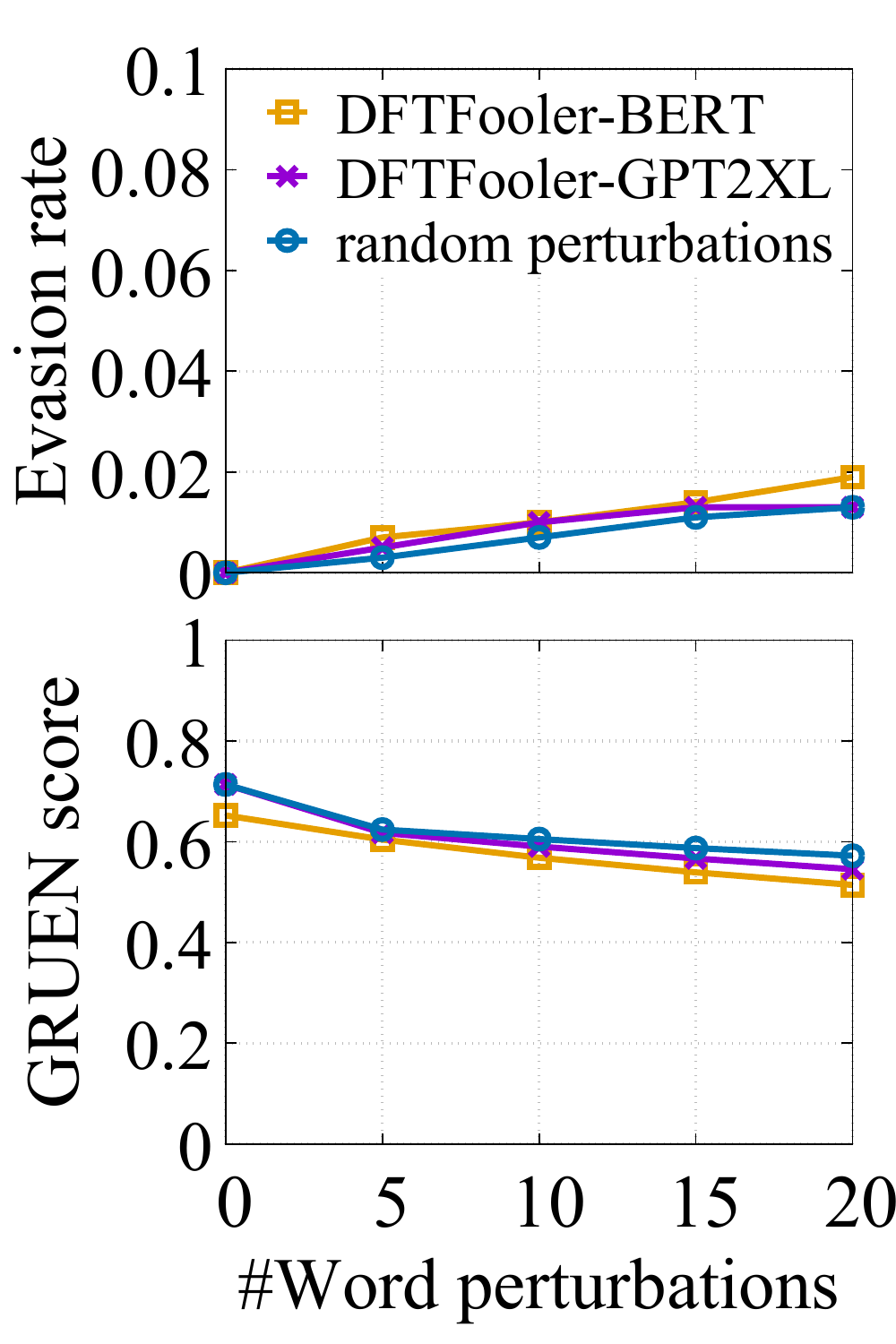}
\caption{Attacking \gbert{}.}
\label{fig:gbert_perturb}
\end{subfigure}
\caption{Evasion rate and average GRUEN score of perturbed text achieved by \ourattack{} and random perturbations when attacking (a) \fast{} and (b) \gbert{}, based on 5, 10, 15, and 20 word perturbations.}
\label{fig:word_perturb_plot}
\end{figure}

\end{document}